\documentclass[final]{jfm}

\usepackage{graphicx}
\usepackage{natbib}
\usepackage{array} 

\ifCUPmtlplainloaded \else
  \checkfont{eurm10}
  \iffontfound
    \IfFileExists{upmath.sty}
      {\typeout{^^JFound AMS Euler Roman fonts on the system,
                   using the 'upmath' package.^^J}%
       \usepackage{upmath}}
      {\typeout{^^JFound AMS Euler Roman fonts on the system, but you
                   dont seem to have the}%
       \typeout{'upmath' package installed. JFM.cls can take advantage
                 of these fonts,^^Jif you use 'upmath' package.^^J}%
      }
  \else
  \fi
\fi

\ifCUPmtlplainloaded \else
  \checkfont{msam10}
  \iffontfound
    \IfFileExists{amssymb.sty}
      {\typeout{^^JFound AMS Symbol fonts on the system, using the
                'amssymb' package.^^J}%
       \usepackage{amssymb}%
       \let\le=\leqslant  \let\leq=\leqslant
       \let\ge=\geqslant  \let\geq=\geqslant
      }{}
  \fi
\fi

\ifCUPmtlplainloaded \else
  \IfFileExists{amsbsy.sty}
    {\typeout{^^JFound the 'amsbsy' package on the system, using it.^^J}%
     \usepackage{amsbsy}}
    {\providecommand\boldsymbol[1]{\mbox{\boldmath $##1$}}}
\fi

\providecommand\bnabla{\boldsymbol{\nabla}}
\providecommand\bcdot{\boldsymbol{\cdot}}

\newcommand\xb{\mathbf{x}}
\newcommand\kb{\mathbf{k}}
\newcommand\Kb{\mathbf{K}}
\newcommand\ub{\mathbf{u}}
\newcommand\kpb{\mathbf{k^{\prime}}}

\newcommand\ir{\mathrm{i}}
\newcommand\er{\mathrm{e}}

\newsavebox{\astrutbox}
\sbox{\astrutbox}{\rule[-5pt]{0pt}{20pt}}

\newcommand\ttz{\ensuremath{\rightarrow 0}}

\title[double-frequency noise generation by surface gravity waves waves]
{Noise generation in the solid Earth, oceans, and atmosphere, from non-linear interacting surface gravity waves in finite depth}
\author[F. Ardhuin and T. H. C. Herbers]%
{F\ls A\ls B\ls R\ls I\ls C\ls E\ns A\ls R\ls D\ls H\ls U\ls I\ls N$^1$\ns \and
T.\ns H.\ns C.\ns H\ls E\ls R\ls B\ls E\ls R\ls S$^2$}

\affiliation{$^1$Ifremer, Laboratoire d'Oc{\'e}anographie Spatiale, Plouzan{\'e}, France \\[\affilskip]
$^2$Department of Oceanography, Naval Postgraduate School, Monterey, California 93943, USA}

\pubyear{2012}
\volume{XX}
\pagerange{1--33}
\date{22 June 2012; revised 22 September 2012; accepted ? }
\begin{document}

\maketitle

\begin{abstract}
Oceanic pressure measurements, even in very deep water, and atmospheric pressure or seismic records, from anywhere on Earth, 
contain noise with dominant periods between 3 and 10 seconds, that is believed to be excited by ocean surface gravity waves. Most of this noise is explained by a nonlinear 
 wave-wave interaction mechanism, 
and takes the form of surface gravity waves, acoustic or seismic waves. Previous theoretical works on seismic noise 
focused on surface (Rayleigh) waves, and did not consider 
finite depth effects on the generating wave kinematics.  
These finite depth effects are introduced here, which requires the consideration of the direct wave-induced pressure at the ocean bottom, a 
contribution previously overlooked in the context of seismic noise.
That contribution can lead to a considerable reduction of the seismic noise source, which  is particularly relevant for  
noise periods larger than 10~s. 
The theory is applied to 
acoustic waves in the atmosphere, extending previous theories that were limited to vertical propagation only. 
Finally, the noise generation theory is also extended  beyond the domain of Rayleigh waves, 
giving the first quantitative expression for sources of seismic body waves. 
In the limit of slow phase speeds in the ocean wave forcing, the known and well-verified gravity wave result is obtained, which was 
previously derived for an incompressible ocean. 
The noise source of acoustic, acoustic-gravity and seismic modes are given by a mode-specific amplification of the 
same wave-induced pressure field near the zero wavenumber. 
\end{abstract}


\section{Introduction}
Ocean waves generate noise in a wide range of acoustic frequencies $f_s$. The upper end of the spectrum, $f_s > 100$~Hz, is dominated by wave breaking and associated bubbles  \citep{Knudsen&al.1948}, whereas the lower 
frequency part, nominally $f_s < 2$~Hz is mostly expected to be caused 
by the nonlinearity of the hydrodynamic equations, on which we focus here. The general sound 
generation by fluid flows was described theoretically  by \cite{Lighthill1952}. 
\cite{Longuet-Higgins1950} showed how seismic waves can be generated by 
the same process, with noise radiating along the Earth's crust in the form of 
Rayleigh waves. That theory was extended to random 
waves by \cite{Hasselmann1963c}, and later cast in the more general framework of wave-wave interactions \citep{Hasselmann1966}. 
Work on compressible flows has also been extended to the study of tsunamis. In that context, \cite{Okal1988} discussed the compressibility 
effect on gravity modes, which we will call here 'acoustic-gravity modes', 
and the gravity effect in seismic 'pseudo-Rayleigh' waves, that we will refer to as 'Rayleigh' modes. 
Interest in seismic noise has risen sharply over the last 
few years with the enforcement of the Comprehensive Nuclear Test Ban Treaty, 
and motivated by the work of  \cite{Shapiro&al.2005} who demonstrated that seismic noise correlation could provide 
a unique monitoring method for the properties of the solid Earth. 

Recent numerical models based on the Longuet-Higgins--Hasselmann (LH-H) theory for 
Rayleigh wave generation have shown good agreement of modeled seismic noise 
spectra with observations \citep{Kedar&al.2008,Ardhuin&al.2011}. It is still 
unclear whether most of the uncertainties on the modeled noise level can be attributed to errors in 
the seismic sources, associated with a poorly constrained
directional distribution of the ocean surface wave spectrum, or to errors in the seismic propagation. 
Seismic observations have also revealed body waves \citep[e.g.][]{Koper&al.2010,Landes&al.2010,Hillers&al.2012}, 
for which no complete theory has been proposed to date.  \cite{Vinnik1973} did propose a  
theory for compressional ($P$) waves, but he did not consider 
the important effect of the water layer. 

Further, the generation of Love waves, 
which are surface shear waves polarized in the horizontal direction, is not well understood.
These Love waves are particularly important
for frequencies below 0.02 Hz or above 1~Hz  \citep[e.g.][]{Bonnefoy-Claudet&al.2006,Nishida&al.2008,Kurrle&Widmer-Schnidrig2008}. 
The low frequency Love waves may be excited by the direct action of long surface gravity waves (known as infragravity waves), on a sloping bottom \citep{Fukao&al.2010}, 
in a way similar to the generation of primary microseisms described by \cite{Hasselmann1963c}.
Here we narrow the scope of our investigation and only consider non-linear wave-wave interactions, which leads to noise with frequencies 
double that of the surface gravity waves.

Finally, the level of acoustic noise has also been explored as a potential source of 
information on the poorly known directional spectrum of short gravity waves \citep{Tyler&al.1974}. \cite{Farrell&Munk2008,Farrell&Munk2010} and \cite{Duennebier&al.2012} showed 
a large variability of the spectral level in the frequency range 0.1--50~Hz that is clearly 
related to the sea state. Their interpretation of the data, following previous studies of 
underwater noise \citep[e.g.][]{Hughes1976,Lloyd1981,Kibblewhite&Ewans1985}, 
is based on sound generated by waves in an unbounded ocean. Although this simplified approach is 
reasonable for high frequency noise, the neglected reflection from the seafloor and subsequent 
reverberations in the bounded ocean may strongly amplify the lower frequency resonant modes.


Given the renewal of interest in seismic and acoustic noise,  
we found it appropriate to revisit Hasselmann's theory. We thus illustrate, correct, and add
a few missing aspects. These corrections include
important terms for intermediate and shallow water that have 
not been considered before for the compressible conditions, although they were 
verified in the incompressible limit \citep{Herbers&Guza1994}.
The compressible equations of motion are used to derive a consistent solution in terms of gravity,
acoustic and seismic modes, including both surface Rayleigh waves and compressional (P) or shear (S) body waves.
This solution for body waves has not been given before.

\begin{figure}
\centerline{\includegraphics[width=0.7\columnwidth]{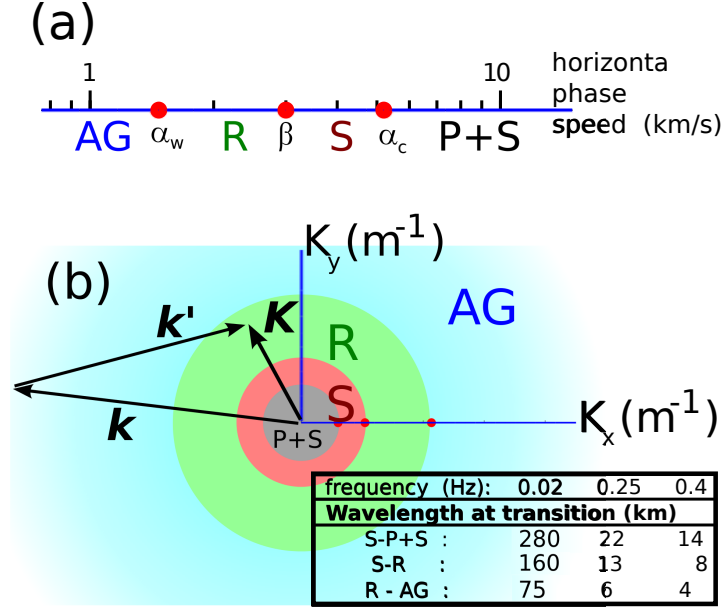}}
  \caption{(a) 
The vertical evanescent or propagating nature of the noise field in the solid and liquid layers is 
defined  by the horizontal phase speeds relative to the distinct values of the sound speed 
in the ocean ($\alpha_w$), and the shear ($\beta$) and compression ($\alpha_c$) speeds in the crust.
 From slow to fast, there are the acoustic-gravity (AG) domain, the Rayleigh (R) wave 
domain, and two body wave domains ($S$ only, and $P$ and $S$ together) 
(b)  For any fixed frequency, the four domains correspond to 4 concentric regions 
in the wavenumber plane. For three selected noise frequencies generated by  OSGW in the  infragravity, dominant and high frequency ranges of the forcing 
wave field, the limiting wavelengths between the four domains are indicated, using $\alpha_w=1.5$~km/s, $\beta=3.2$~km/s, 
$\alpha_c=5.54$~km/s. One example of interaction (black vectors) is shown with 
two gravity wave modes that interact to generate a Rayleigh wave.}
\label{fig:4domains}
\end{figure}
Building on the seismo-acoustic paradigm proposed by \cite{Arrowsmith&al.2010}, the basic idea of the present paper is that all modes of 
motion can be excited by ocean surface gravity waves (OSGW) of any frequency. 
For a given pair of interacting frequencies $f$ and $f'$, the frequency of the generated noise is $f+f'$, and the different types of waves 
are only distinguished by their phase velocity, or equivalently by their horizontal wavenumber $K$ which is the norm of the vector 
sum of the wave numbers, $\Kb= \kb +\kpb$ of the interacting OSWG, as shown in figure \ref{fig:4domains}. 
This type of wave-wave interaction is one of the lowest order interactions \citep{Hasselmann1966}. 

\begin{figure}
\centerline{\includegraphics[width=\columnwidth]{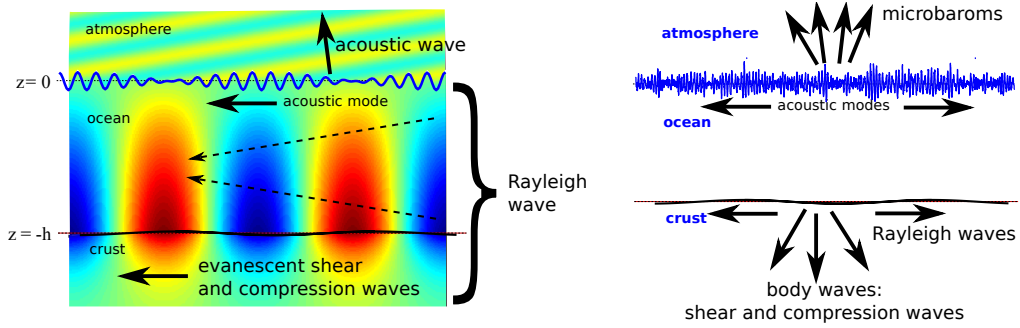}}
  \caption{Left, schematic of second-order pressure field, in colours, associated with double-frequency noise, 
forced by the interaction of a single pair of directionally opposing monochromatic wave trains on the sea surface that form periodic groups.
The individual waves in the group advance very slowly, at the OSWG phase speed.
However, in this particular case, the group as a whole travels slightly faster than the sound speed in water, and slower than the crust 
shear and compression velocity. As a result, the crust elastic waves are evanescent while ocean and atmospheric waves propagate also 
in the vertical. In the ocean, the superposition of upward and downward waves (dashed lines) gives the 
vertical mode structure. The number 
of waves in the group was reduced for visibility, and for the same reason 
the amplitudes of the sea surface and bottom elevations have been exaggerated as well as the pressure fluctuation 
in the atmosphere relative to those in the ocean. 
Right, schematic of ocean waves with a relatively broad spectrum, giving rise to the interaction of all possible pairs of wave trains
 and  noise radiation in all directions.}
\label{fig:pressure_field}
\end{figure}

In the physical space, one such interaction excites waves with horizontal wavelength $L_h=2 \pi/K$, and different 
vertical patterns in the atmosphere, ocean and crust, due to the very different speeds for compression waves in these three media. Figure \ref{fig:pressure_field}
shows how the same forcing can give almost vertical-propagating waves in the atmosphere, waves that propagate almost horizontally in the ocean, 
and evanescent waves in the crust. The analysis of a pair of interacting wave-trains is 
key to our interpretation of the different noise modes.  The broad-banded wave spectrum of ocean waves results in the superposition 
of all possible pairs of OSGW waves trains, and thus all possible noise waves propagating or 
evanescent, that radiate in all directions.  This broad spectrum allowed \cite{Longuet-Higgins1950} to consider all interactions at once, 
replacing the wave-induced forcing by an equivalent point source exerted on the sea surface. However, that latter approach is only valid 
for noise wavelengths much larger than those of the interacting OSWG.
Table 1 points directly to the main theoretical results of the present paper, 
and all symbols and notations are listed in tables \ref{table_symb}--\ref{table_symb3}.
 \begin{table}
  \centering
 \caption{Summary of theoretical expressions for the gravity, acoustic and seismic noise fields. 
 \label{tablesummary}}
  \begin{tabular}{lcc}
                      & infinite water depth           & finite water depth $h$ \\
Surface forcing field &   eq. (\ref{eq:Fp3D})         &  eq. (\ref{p2_spectrum})        \\
Bottom  forcing field &   n.a.                        &  eq. (\ref{P2bot})        \\
gravity-like modes    &   eq. (\ref{Fp1Dag})          &   not given       \\
 acoustic spectrum    &    eq. (\ref{eq:noise_Lloyd}) &  eq. (\ref{eq:sound_finite})  \\
microbaroms           &    eq. (\ref{Fp1Dap})         &   eq. (\ref{Fp1Dap})          \\
 seismic source (R)    &    n.a.                       &  eq. (\ref{SDF_k})      \\ 
seismic source (P)    &    n.a.                        & eq. (\ref{eq:P_source})  \\
seismic source (S)    &    n.a.                       &  eq. (\ref{eq:S_source})  \\
seismic spectrum (R)  &    n.a.                        &  eq. (\ref{F_delta})                       \\
seismic spectrum (P)  &   n.a.                        &   eq. (\ref{F_delta_P})                      \\
\end{tabular}
\end{table}

\section{Water waves theory and noise sources}
Here we give only a brief derivation of the solution, which is a straightforward compressible extension to the 
solution given by \cite{Hasselmann1962}. Water column motions are expanded in powers of the sea surface 
slope with a linear motion for which compressible effects may be neglected \citep{Longuet-Higgins1950}, and a second order motion 
with pressure $p_2$ and velocity potential $\phi_2$. It is for this second order motion that we are extending previous results. 
Following again  \cite{Longuet-Higgins1950} we will see that $\phi_2$ approximately obeys the same linear acoustic wave equation 
as $\phi_1$, so that the forcing of the double-frequency noise is only due to boundary conditions. 
In contrast to  \cite{Hasselmann1963c}, we consider here the general finite depth expression 
for the surface pressure forcing $\widehat{p}_{2,\mathrm{surf}}$, and in the boundary condition 
at the ocean bottom the additional forcing $\widehat{p}_{2,\mathrm{bot}}$ that accounts for the Bernoulli 
effect of the near-bed orbital wave motion. 
The theory presented here extends the second order finite depth theory of \cite{Hasselmann1962} to a compressible 
ocean and elastic seafloor.

\subsection{Equations of motion}

We decompose the water density into a mean value
$\rho_w$ and a perturbation $\rho \ll \rho_w$. 
Neglecting stratification effects due to temperature and salinity, 
the fluctuations in pressure $p$ and water density $\rho$ are related by an equation of state, 
which involves the speed of sound in water $\alpha_w$ \citep[][eq. 32]{Lighthill1978}.
\begin{eqnarray}
  \frac{d p}{d t}= \alpha_w^2 \frac{d \rho}{d t} \label{eq:state},
\end{eqnarray}

We assume that the motion is irrotational 
so that the velocity field is given by the gradient of 
the velocity potential $\phi$. 
This assumption, which implies a frictionless interior ocean, is well supported by the observed weak attenuation of 
swells propagating across ocean basins \citep{Ardhuin&al.2009b} and local comparisons of the observed wave orbital motion with second order wave theory
\citep[e.g.][]{Herbers&al.1992}. Vorticity effects can be important in the vicinity of the surface and bottom boundary 
layers \citep[e.g.][]{Longuet-Higgins1960}, or in the presence of sheared currents \citep[e.g.][]{Peregrine1976}. The former effect has little influence on the pressure field, which is our primary 
interest, and we will not consider here the effects of currents. 
Because of the importance of the apparent
gravity acceleration $g$, which defines the vertical axis, we separate 
horizontal and vertical components
using vectors and gradient operators in the horizontal plane, e.g.
$\ub=\bnabla \phi=(\partial \phi/\partial x,\partial \phi/ \partial y)$ and $w = \partial \phi / \partial z$.
The conservation of mass of sea water is 
\begin{eqnarray}
\frac{d \rho}{d t}&=& - \rho_w \nabla^2 \phi - \rho_w \frac{\partial^2 \phi}{\partial z^2}  \label{eq:sismo_rho_w}.
\end{eqnarray}

Equations (\ref{eq:sismo_rho_w})--(\ref{eq:state}) can be combined to eliminate $\rho$, 
\begin{eqnarray}
\frac{d p}{d t}=   -\rho_w \alpha_w^2 \left[ 
\nabla^2 + \frac{\partial^2 }{\partial z^2}\right] \phi \label{eq:sismo_p_w2a}
\end{eqnarray}

From eq. (\ref{eq:state}), the water density is only a function of the pressure. 
The two unknowns $p$ and $\phi$ are also related by the momentum conservation equation, 
with can be cast in the form of Bernoulli's equation 
 \citep[see e.g.][section 20]{Lamb1932},
\begin{equation}
    \frac{\partial \phi}{\partial t} = 
    -\frac{1}{2}\left[
    \left|\bnabla \phi\right|^2
    +\left(\frac{\partial \phi}{\partial z}\right)^2
    \right]-\frac{p}{\rho_w} - gz + C(t),
\label{Bernoulli2}
\end{equation}
with $C(t)$ a time-varying but spatially uniform function.

The boundary conditions at the surface $z=\zeta$ are given by the continuity of pressure and vertical velocity
\begin{eqnarray}
    p&=&p_a   \label{psurf} \\
      \frac{\partial \phi }{\partial z}&=& \bnabla {\mathbf \phi}\bcdot\bnabla \zeta
    + \frac{\partial \zeta }{\partial t} \quad{\mathrm{at}} \quad \quad z=\zeta \label{wsurf},
\end{eqnarray}
with the atmospheric pressure $p_a$. This expression is translated to the mean sea level $z=0$ with a  Taylor expansion for $\phi$
\begin{equation}
     \frac{\partial \zeta }{\partial t} - \frac{\partial \phi }{\partial z} \simeq 
- \bnabla {\mathbf \phi}\bcdot\bnabla \zeta  + \zeta \frac{\partial^2 \phi }{\partial^2 z} \quad{\mathrm{at}} \quad z=0. \label{surf_kine_Taylor}
\end{equation}

Following \cite{Longuet-Higgins1950}, we shall now expand the solution in powers of the surface slope, with 
 the sea surface elevation $\zeta_1$, associated with linear waves, and a 
non-linear correction $\zeta_2$ such that $|\zeta_2| \ll |\zeta_1|$ \citep[see also e.g.][]{Hasselmann1962}, 
\begin{subeqnarray}
\gdef\thesubequation{\theequation \textit{a,b}}
    \zeta=\zeta_1 + \zeta_2, \quad \phi=\phi_1 + \phi_2. 
\end{subeqnarray}

\subsection{Linear solution}
We consider the case of a constant depth $h$. 
Compressibility effects in the linear solution  are negligible for our purpose \citep[see][eq. 123]{Longuet-Higgins1950}, 
so that we may use 
\begin{subeqnarray}
\gdef\thesubequation{\theequation \textit{a,b}}
\zeta_1=\sum_{\kb,s} Z_{1,\kb}^{s} \mathrm{e}^{\mathrm{i}
     (\kb \bcdot \xb - s \sigma t) } \label{eq:zeta1}, \quad  \phi_1 \simeq \sum_{\kb,s}  -\ir \frac{s g}{\sigma}  \frac{\cosh(kz+kh)}{\cosh(kh)}    Z_{1,\kb}^{s}
 \mathrm{e}^{\mathrm{i}  
     (\kb \bcdot \xb - s \sigma t) }  \label{eq:phi1}
\end{subeqnarray}
where $k$ is the norm of the horizontal wavenumber vector $\kb$, $s$ is a sign index equal to -1 or 1, 
so that $s=1$ corresponds to waves propagating in the direction of the vector $\kb$, and $s=-1$ corresponds to the opposite direction. 
The radian frequency $\sigma$ is given by the dispersion relation for linear waves \citep{Laplace1776} , 
\begin{equation}
 \sigma  = \sqrt{g k \tanh (kh)} \label{eq:disp_Airy},
\end{equation}
giving the group speed 
\begin{equation}
 C_g = \frac{\partial \sigma}{\partial k}  = \frac{\sigma}{k}\left[\frac{1}{2} + \frac{kh}{\sinh(2kh)}\right] \label{eq:Cg}.
\end{equation}

\subsection{The negligible near-surface forcing}
Eliminating $p$ between
(\ref{eq:sismo_p_w2a}) and (\ref{Bernoulli2}), we obtain the acoustic wave equation
\begin{equation}
\alpha_w^2 \left[ \nabla^2 + \frac{\partial^2 }{\partial z^2}\right] \phi 
= \frac{d}{dt}\left[\frac{\partial \phi}{\partial t} +\frac{1}{2}\left(\left|\bnabla \phi\right|^2
 + \left(\frac{\partial \phi}{\partial z}\right)^2\right)\right]  \label{eq:sismo:forcing2a},
\end{equation}
where the gravity term has been removed 
by our approximation of a constant mean density. That term is usually neglected 
eventually in the solution \citep{Stoneley1926}. The expression for the $C(t)$ term in eq.  (\ref{Bernoulli2}), 
is given by eq. (28) in \cite{Longuet-Higgins1950} and it is actually zero provided 
that there are no standing waves, i.e. we do not consider the case of waves trains of 
exactly equal frequency and opposite direction. For broad wave spectra we may indeed neglect these contributions
since the measure of such pairs of wave components is zero while the measure of \emph{nearly} 
opposite waves is finite. 

Our eq. (\ref{eq:sismo:forcing2a}) corresponds to eq. (130) in \cite{Longuet-Higgins1950}, who showed that the non-linear terms 
yield a contribution (his $F$ terms) that is of the order of  $g/(2 \alpha_w^2 k)$ times  
the other terms. Even for very long waves with a wavelength of 50~km, this factor 
is 0.02, so that we shall neglect this source of wave forcing. This leads to a 
linear wave equation for the second order velocity potential 
\begin{equation}
 \frac{\partial^2 \phi_2}{\partial^2 t}  \simeq  \alpha_w^2 \left[ \nabla^2 + \frac{\partial^2 }{\partial z^2}\right] \phi_2, \quad \mathrm{for} \quad -h\leq z \leq 0 \nonumber \\
\label{eq:sismo:forcing2lin}
\end{equation}

\subsection{Surface forcing}
Because the compressibility only affects the mass conservation equation, it does not modify the kinematic and dynamic boundary conditions. 
These are given by \cite{Hasselmann1962}. The unknown $\zeta$ is eliminated from the linear terms by adding $\partial$(\ref{Bernoulli2})$/\partial t$ evaluated with (\ref{psurf}), 
and $g \times $~(\ref{surf_kine_Taylor}). This combination of kinetic and dynamic boundary condition 
gives an equation for the velocity potential to second order in the wave slope, valid at $z=0$. Keeping the lowest 
order non-linear terms we have, 
\begin{equation}
\left(\frac{\partial ^{2}}{\partial t^{2}}
    + g \frac{\partial }{\partial z}\right) \phi_2
     = - g \left[ \bnabla  \phi_1 \bcdot  \bnabla \zeta_1  + \zeta_1  \frac{\partial^2 \phi_1 }{\partial^2 z} \right] 
- \frac{1}{2} \frac{\partial }{\partial t}\left[
\left|\bnabla \phi_1 \right|^2     +\left(\frac{\partial \phi_1 }{\partial z}\right)^2 +2 \zeta_1  \frac{\partial^2 \phi_1 }{\partial t \partial z}\right].
\label{eq:Bernoulli2z0}
\end{equation}

We now give the explicit form of the right hand side using eqs. (\ref{eq:zeta1}). 
As noted by \cite{Hasselmann1963c}, we may rewrite (\ref{eq:Bernoulli2z0}) as 
\begin{equation}
\left(\frac{\partial ^{2}}{\partial t^{2}}
    +g \frac{\partial }{\partial z}\right) \phi_2 = -\frac{1}{\rho_w}\frac{\partial \widehat{p}_{2,\mathrm{surf}} }{\partial t} \label{eq:eq_phi_2}.
\end{equation}
Namely, our problem is equivalent to the effect of a pressure field  
$\widehat{p}_{2,\mathrm{surf}}$ applied at $z=0$. Using the linear solution (\ref{eq:phi1}),
\begin{eqnarray}
     \widehat{p}_{2,\mathrm{surf}}= \rho_w \sum_{\kb,s,\kpb,s'}  D_z \left(\kb,s,\kpb,s'\right) Z_{1,\kb}^{s} Z_{1,\kpb}^{s'} 
\mathrm{e}^{\mathrm{i}
    \Theta(\kb,\kpb,s,s')} \nonumber \\ \label{P2hat}
\end{eqnarray}
with the phase function of interacting wave trains defined by
\begin{equation}
\Theta(\kb,\kpb,s,s')=\left[\left(\kb+\kpb\right)\bcdot{\mathbf x} - \left(s \sigma+s' \sigma'   \right) t\right]
\end{equation}
and coupling coefficient $D_z \left(\kb,s,\kpb,s'\right)$  given by eq. (\ref{Dz}).

\begin{figure}
\centerline{\includegraphics[width=0.8\columnwidth]{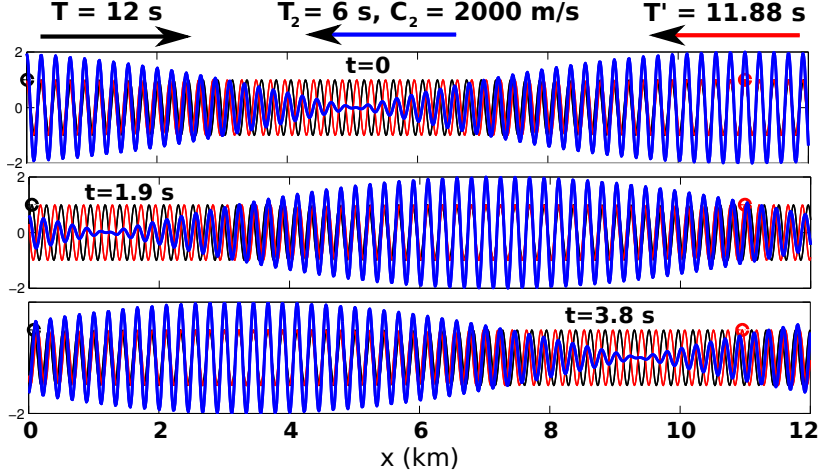}}
  \caption{Generation of supersonic wave groups (in blue) by the superposition of 
two opposing deep water monochromatic wave trains of nearly equal periods $T$ and $T'$. 
The curves show the surface elevation of the individual wave trains (black and red) or their combination (blue). 
The group with length 12~km propagates in the same direction as the wave train with the shortest 
wavelength. If the wave trains are not exactly directionally opposing, the group propagates in the direction of the vector $\Kb= \kb +\kpb$. 
The red and black dots 
are attached to the wave crest of each train, and move 100 times slower than the blue group.}
\label{fig:group_negative}
\end{figure}

The shape of this pressure pattern is well understood when the full sum in eq. (\ref{P2hat}) is reduced to only two interacting 
deep water wave trains with amplitudes $a$ and $a'$ and 
slightly different frequencies $\sigma$ and $\sigma'$, 
traveling in opposite directions, as shown in figure \ref{fig:group_negative}.  For  $kh \gg 1$ we may use
$D_z\left(\kb,1,-\kb,1\right)= - 2 \sigma^2$ and 
$D_z\left(\kb,1,\kb,1\right)=  0$.
Defining 
$K=k-k'$,  the 
equivalent surface pressure is given by the sum interaction, 
\begin{equation}
     \widehat{p}_{2,\mathrm{surf}} = -2 \rho_w \sigma^2 aa'
     \left\{ \cos\left[K  x + 2  \sigma   t\right] \right\} \label{group}. 
\end{equation}
 The corresponding difference interaction yields a short wave, with wavenumber $2 k$, 
that will not be considered here \citep[see e.g.][]{Hasselmann1963c}.
We note that the wave forcing $\widehat{p}_{2,\mathrm{surf}}$ is out of phase 
with $-u^2/2$, the part of the Bernoulli head which comes from the horizontal velocities. Indeed, 
as the wave trains propagate in opposite directions their velocity partially cancel 
where the surface elevations add up, thus causing a pattern of higher pressure under the groups of high waves 
and lower pressure under the lulls, that is opposite to the familiar Bernoulli effect of set-down under groups 
of uni-directional waves \citep{Longuet-Higgins&Stewart1962}.  
As a result the double-frequency perturbations of  $-u^2/2$ have a sign 
opposite to $\widehat{p}_{2,\mathrm{surf}}$. That effect will be very important for waves in shallow water. 
In deep water, the contribution of $-w^2/2$ must also be considered. Using $\sigma \simeq \sigma'$, we have
\begin{eqnarray}
     u &=& \sigma a \cos(kx -\sigma t)  - \sigma' a' \cos(kx +\sigma t) \\
     w &=& \sigma a \sin(kx -\sigma t)  - \sigma' a' \sin(kx +\sigma t) \\
     -(u^2 + w^2)/2 &=&  aa' \sigma^2  \cos\left[K  x + 2  \sigma   t\right] - \sigma^2 \left(a^2  + a'^2 \right)/2.
\end{eqnarray}

Eq. (\ref{group}) generalizes the result given by \cite{Longuet-Higgins1950} for 
equal wave periods. The standing wave studied by \cite{Longuet-Higgins1950} is thus obtained, 
somehow paradoxically, as the limit of wave groups that travel  at an 
infinite speed, but that are infinitely long.
 
For wave directions nearly opposite, instead of exactly opposite, this first term can 
propagate in any horizontal direction, given by the direction of $\Kb=\kb+\kpb$. 

\subsection{Noise spectrum and finite depth effects}
Both LH50 and H63 used the deep-water approximation 
\begin{equation}
\widehat{p}_{2,\mathrm{surf}} \simeq \rho_w \left[ \left|\bnabla \phi\right|^2
    +\left(\frac{\partial \phi}{\partial z}\right)^2  \right] \quad \mathrm{valid} \quad \mathrm{for} \quad kh \gg 1, \label{p2_deep}
\end{equation}
instead of the more complex but more general form given by eq. (\ref{P2hat}). 
We will illustrate the important differences between these two expressions for $kh < 1$, by considering the 
forcing of very long components. Indeed, sound waves in the ocean have velocities in excess of 1.4~km~s$^{-1}$. 
As a result, regardless of the vertical wavenumber $l$, the horizontal wavenumber vector is relatively small $K = \sqrt( \omega_s^2/ \alpha_w^2 - l^2) < \omega_s/\alpha_w$. 
Thus the acoustically noisy wave interactions verify $K \ll k$,
 which gives $\kpb \simeq -\kb$, and $f\simeq f'$ with $f_s \simeq 2 f$.
We may thus focus on the estimation of the spectrum of $\widehat{p}_{2,\mathrm{surf}}$ at $\Kb \simeq 0$. 

For this purpose we introduce the spectral density of this surface pressure 
in the three spectral dimensions $(K_{x},K_{y},f_s)$. Using the  Fourier amplitude $\widehat{p}_{2,\mathrm{surf}}(\Kb,f_s)$ of the forcing pressure 
$\widehat{p}_{2,\mathrm{surf}}$, with wavenumber vector $\Kb$ 
and frequency $f_s$, 
\begin{equation}
F_{p2,\mathrm{surf}}(\Kb,f_s) = 2 \lim_{\left|d \Kb\right| \ttz, df_s \ttz} \frac{\left|\widehat{p}_{2,\mathrm{surf}}(\Kb,f_s) \right|^2}
{\mathrm{d}K_{x} \mathrm{d}K_{y} \mathrm{d} f_s},\label{Fp3D_def} 
\end{equation}
The factor 2 in the expression makes this a single-sided spectrum, with non-zero values only for $f_s \geq 0$. 
This spectral density is in a three-dimensional spectral space, with S.I. units of Pa$^2$~m$^2$~s; it is denoted $F_{p3D}$ in \cite{Ardhuin&al.2011}. 

Using (\ref{P2hat}) the surface pressure spectrum can be expressed in terms of quadratic products of the 
(linear) sea surface elevation spectrum   
\begin{equation}
E(k_x,k_y)  = 2 \lim_{\left|d \kb \right| \ttz } \frac{\left|Z_{1,\kb}^{+}\right|^2}{\mathrm{d}k_x \mathrm{d}k_y}
\end{equation}
with a coupling coefficient from (\ref{Dz}) that simplifies for $K \simeq 0$ to   
\begin{equation}
   D_z\left(\kb,1,-\kb,1\right)   =  - 2 \sigma^2\left[1 + \frac{1}{4 \sinh^2(kh)}\right]. \label{Dz_keq0}
\end{equation}

To transform the spectra to frequency-direction space we use the Jacobian transformation  
\begin{equation}
 E(f,\theta) =\frac{2 \pi k}{ C_g} E(k_x,k_y). 
\end{equation}
We now introduce the directional distribution  $M$ such that  $E(f,\theta)=E(f) M(f,\theta) $,  and we define 
the directional integral 
\begin{equation}
I(f)=\int_{0}^{\pi}M(f,\theta)M(f,\theta+\pi)\mathrm{d}\theta.\label{eq:I}
\end{equation}
With these notations we finally obtain 
\begin{eqnarray}
     F_{p2,\mathrm{surf}}(\Kb \simeq 0,f_s)   &\simeq&   \rho_w^2  D_z^2 \int  E(f,\theta) E(f,\theta+\pi) 
\frac{C_g^2 \mathrm{d}k_x \mathrm{d}k_y }{k^2 4 \pi^2 \mathrm{d} f_s}   \nonumber\\
&\simeq &   \rho_w^2 g^2  f_s E^2(f)I(f)  \tanh^2(kh)  \left[1+\frac{2 kh}{\sinh(2kh)} \right]  
\left[1+ \frac{1}{4 \sinh^2(kh)} \right]^2   \nonumber\\
\label{p2_spectrum}
\end{eqnarray}
In deep water ($kh \gg 1$), the equivalent surface pressure is $\widehat{p}_{2,\mathrm{surf}} \simeq \rho_w (u_1^2+w_1^2)$, 
and its spectrum  \citep{Hasselmann1963c,Ardhuin&al.2011}
\begin{equation}
F_{p2,\mathrm{surf}}(\Kb \simeq 0,f_s)=\rho_w^2 g^2  f_s   E^2(f)I(f).\label{eq:Fp3D}
\end{equation}

As shown in figure \ref{fig:finitedepth}, this spectral density of surface pressure is, in finite depth, 
up to four times smaller than the deep water approximation obtained 
from eq. (\ref{p2_deep}) and used by  \cite{Webb2007} and \cite{Tanimoto2006}. 
This is because the surface pressure actually combines two terms. One is $ \rho_w (u_1^2+w_1^2)/2$, from the momentum 
conservation equation -- which here takes the form of the Bernoulli equation (\ref{Bernoulli2}) -- and the other, 
equal in magnitude in deep water,  comes from the non-linearity of 
the surface boundary condition (\ref{surf_kine_Taylor}). As $kh$ goes to zero, both 
the latter term and the 
vertical velocity contribution to the Bernoulli pressure are small 
compared with $\rho_w u_1^2/2$, thus reducing the surface pressure 
forcing by a factor four.

This effect is not important 
for the dominant microseismic peak, generated by waves of period about 10~s, but it may be important for 
the much longer waves, known as hum, driven by long surface gravity waves. However, in that case, one should also consider
the direct action of the wave-induced pressure on the bottom.

\subsection{Additional bottom forcing}
To evaluate noise in the water column, 
both the second order pressure forcing applied at the 
surface and bottom have to be taken into account. This includes 
the well known Bernoulli effect of a pressure drop in response to an increase in velocity. 
At the bottom, this wave-induced pressure is $-\rho_w u_1^2/2$ which is  exactly out of phase with  $\widehat{p}_{2,\mathrm{surf}}$. 
 In the limit $kh \ttz$, it cancels the source of noise that would have resulted from the surface forcing alone.
In physical terms, for $kh=0$, we have the same momentum balance at the sea surface and bottom, and thus 
the same pressure perturbations, which are zero because the pressure just below the surface has to match the atmospheric pressure.

For any value of $kh$, 
the coupling coefficient given by eq. (\ref{Dz_keq0}) differs from the full second order coefficient for the bottom pressure
\citep[e.g. eq. 4 in][]{Herbers&Guza1991}, 
which also involves the Bernoulli head (the bracket in eq. \ref{Bernoulli2}). However, that extra term 
is also relevant to the generation of seismic noise due to the  
bottom boundary condition that couples the solid crust to the water column. Indeed, the second-order 
pressure perturbation at the bottom writes, 
\begin{equation}
p_2(-h)=- \rho_w \frac{\partial \phi_2}{\partial t} + \widehat{p}_{2,\mathrm{bot}}, \label{pb}
\end{equation}
where the Bernoulli head contribution to the pressure can be expressed from the first order wave amplitudes, 
\begin{equation}
\widehat{p}_{2,\mathrm{bot}}= \rho_w \sum_{\kb,s,\kpb,s'}  D_{pb} \left(\kb,s,\kpb,s',z=-h\right) Z_{1,\kb}^{s} Z_{1,\kpb}^{s'} 
 \mathrm{e}^{\mathrm{i}
     \Theta(\kb,\kpb,s,s')}\label{P2bot},
\end{equation}
with a coupling coefficient $D_{pb}$ given by eq. (\ref{Dpb}).

We may interpret the bottom pressure (\ref{pb}) as the sum of the surface forcing $\widehat{p}_{2,\mathrm{surf}}$ transmitted 
to the bottom by  $\phi_2$, and  a direct effect of the Bernoulli head at the bottom which 
is an additional forcing $\widehat{p}_{2,\mathrm{bot}}$ that partly cancels $\widehat{p}_{2,\mathrm{surf}}$. 

\begin{figure}
\centerline{\includegraphics[width=0.8\columnwidth]{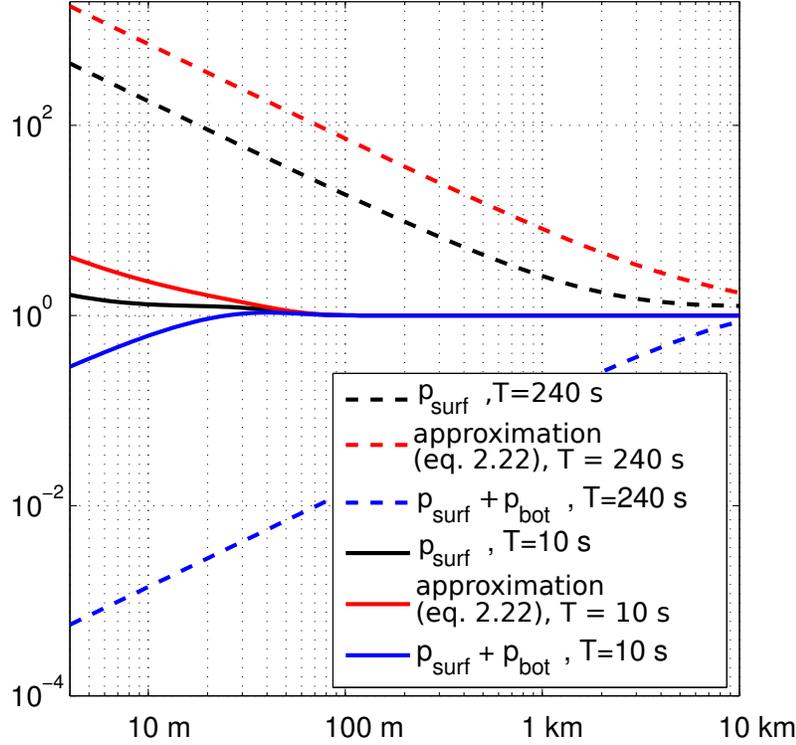}}
  \caption{Compared to deep water, the source of seismic noise power in finite water depth is 
amplified by a factor 
$\tanh^2(kh)  \left[1+{2 kh}/{\sinh(2kh)} \right]$, when accounting for both the surface 
and bottom forcing. This is very different from the approximation that considers the surface forcing only (leading to eq. \ref{p2_spectrum}), 
even more so when using its deep water approximation  (eq. \ref{p2_deep}) which gives an additional factor 4 difference for $kh \ll 1$.
\label{fig:finitedepth}}
\end{figure}

We shall see in the next section that the forcing term for seismic noise 
is $\widehat{p}_{2,\mathrm{surf}} +\cos(lh) \widehat{p}_{2,\mathrm{bot}}$, with $l \leq K \ll k$ 
the vertical wavenumber in the water.  For shallow water gravity waves, $kh \ll 1$ and thus $\cos(lh) \simeq 1$ so that 
the effective forcing term becomes $\widehat{p}_{2,\mathrm{surf}} + \widehat{p}_{2,\mathrm{bot}}$, 
which equals the bottom pressure in the  incompressible limit. 
The shallow water asymptote of the spectrum of this total forcing term is 
very different from the surface pressure only. Compared to eq. (\ref{p2_spectrum}), 
the  $\left[1  + 0.25 / \sinh^2(kh)\right]^2$ factor is now replaced by 
$1$. For $kh \ll 1$, this is a factor $(kh)^4/16$ smaller, as shown in figure \ref{fig:finitedepth}. 
This asymptote is relevant for the hum, the noise with periods larger than 30~s, which is believed to be driven 
by long (infragravity) surface gravity waves \citep{Webb2007}.
The source of this hum is attenuated by several orders of magnitude   on the continental shelves
and not amplified according to the deep water approximation given by  eq. (\ref{p2_deep}). 

\section{From surface pressure spectrum to noise spectra}
\subsection{Dispersion relation and modes}
The problem of noise generation by waves has been reduced 
to that of noise generation by an equivalent surface pressure field
$\widehat{p}_{2,\mathrm{surf}}(x,y,t)$, with the possible addition of a bottom pressure field 
$\widehat{p}_{2,\mathrm{bot}}(x,y,t)$ for finite depths. From a statistical point of 
view, this equivalent surface pressure is fully represented by its spectrum, 
$F_{p2,\mathrm{surf}}(\Kb,f_s)$ where $\Kb=\kb+\kpb$ is the sum of the 
interacting wavenumbers and $f_s=f\pm f'$ is the sum or difference of 
the interacting frequencies. Subjected to this surface forcing, our linearized 
wave equation (\ref{eq:sismo:forcing2lin}) will have linear solutions. 
In particular, any propagating or evanescent solution will take the following form
\begin{equation}
 \phi_2 \propto \exp[\ir(K_{x} x +K_{y} y + l z - \omega_s t)],
\end{equation} 
which, substituted in  the wave equation, gives the dispersion relation 
\begin{equation}
 \left\{-\omega_s^2  + \alpha_w^2 \left[ K^2 + l^2 \right]
 \right\}= 0, \label{eq:sismo:forcing2lin_fourier}
\end{equation}
where $\omega_s=2 \pi f_s$. Both $\omega_s$ and $K=\left|\Kb\right|$ are imposed by the forcing, 
so that the magnitude of the complex vertical wavenumber $l$ is given by
\begin{equation}
 l=  K \sqrt{\frac{\omega_s^2}{K^2 \alpha_w^2} - 1},
\end{equation}
which yields
\begin{subeqnarray}
 \gdef\thesubequation{\theequation \textit{a,b}}
\phi_2 =  \left ( C \er^{il(z+h)} + D \er^{-il(z+h)}\right) \er^{\ir \Theta}, \quad \Theta= (K_{x} x +K_{y} y  - \omega_s t) \label{phi2CD}
\end{subeqnarray}
where $C$ and $D$ are the bottom amplitudes 
of the upward and downward propagating waves, determined by the surface and bottom boundary conditions. 
\begin{figure}
\centerline{\includegraphics[width=\columnwidth]{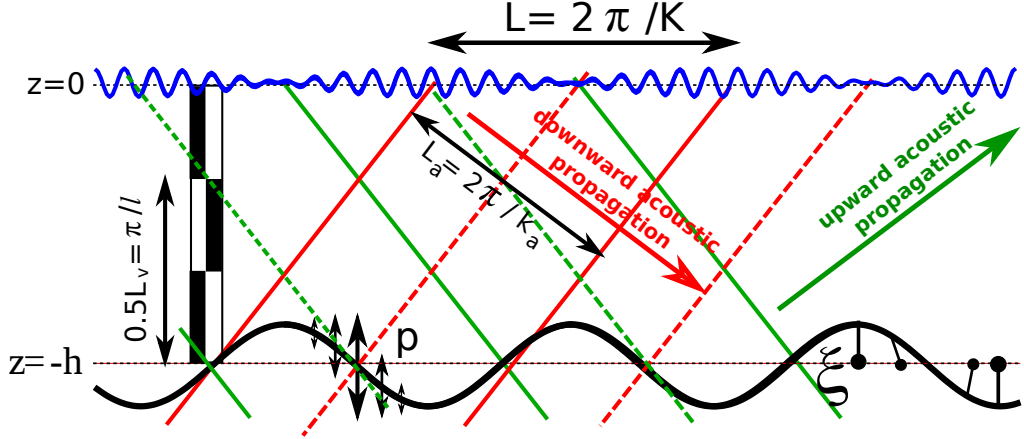}}
  \caption{Schematic of wave groups and forced acoustic and seismic wave motion. For a given wave group period $T$, 
the horizontal wavelength $L=2\pi /K$ 
can be larger than the acoustic wavelength in the water due to the oblique incidence of the sound waves.  
The superposition of two obliquely 
propagating sound waves (arrows) forms a mode pattern that propagates horizontally at a supersonic speed. 
The acoustic wavelength is $L_a=2\pi /k_a=2\pi /\sqrt{K^2+l^2}=\alpha_w/T$. Both vertical and horizontal wavelengths 
are larger. 
For readability, the wave and bottom amplitude is not to scale, and we have reduced the number of 
waves in the group from 104  to 10. Other than that, the angles are preserved. 
The configuration shown here corresponds to the conditions for maximum amplification of mode 1 (see below), with a vertical 
wavelength to water depth ratio of 0.75. For a water depth $h=4400$~m depth, this corresponds to 
$L=7.7$~km and a seismic frequency $f_s=0.29$~Hz. }
\label{fig:water_and_crust}
\end{figure}


There are two classes of solutions. Those for which  $\omega_s/K \le \alpha_w$  and thus $l$ is imaginary: these are 'acoustic-gravity
' modes with an 
amplitude that decays exponentially from the surface. For the shortest components, we have $\omega_s/K \ll \alpha_w$ 
and thus $l= \ir k$, corresponding 
to the incompressible limit in which only gravity is important. 
The other class of  solutions, for which  $\omega_s/K \ge \alpha_w$ and  $l$ is real, are the acoustic modes that propagate along both vertical and horizontal 
dimensions, as illustrated in figure \ref{fig:water_and_crust}. 

The equations of elasticity in the crust yield the same wave equation, one 
for compressional and another for shear motions, with $\alpha_w$ replaced by the crusts compressional wave speed
$\alpha_c$ and shear wave speed $\beta_s$, respectively \citep[e.g.][]{Aki&Richards2002}. 
Since we have $\alpha_w < \beta_s < \alpha_c$, we can distinguish four different 
regimes depending on the value of the horizontal phase speed $\omega_s/K$, relative to these three velocities (fig. \ref{fig:4domains}).

In the family of gravity-acoustic modes, there is no resonance in the forcing, namely 
there are no free waves with $\Kb=\kb+\kpb$ and $f_s=f\pm f'$ and thus the motion is locked with the forcing 
wave groups, as verified by \cite{Herbers&Guza1994}.  The acoustic modes propagate obliquely to the vertical 
away from the surface, and interact with the ocean bottom (figure \ref{fig:water_and_crust}). The seismic Rayleigh waves modes 
propagate horizontally, combining an acoustic-like motion in the water and evanescent elastic waves in the crust, illustrated 
by figure \ref{fig:pressure_field}. 
The coupling with the bottom motion selects 
a few resonant modes that dominate the solution, with an energy that grows linearly with the propagation distance 
\citep{Hasselmann1963c}. For wavenumbers 
that allow a vertical propagation in the crust, we obtain body waves. These waves can be compression or shear ($P$ or $S$) waves, 
and these two types occur in overlapping ranges of $K$.

\subsection{Acoustic noise in an unbounded ocean}
In order to illustrate the different types of solutions, it is interesting to evaluate the solution for an 
unbounded ocean, 
in which sound waves are radiated from the surface only. 
The velocity field and associated pressure fluctuations are 
\begin{eqnarray}
\phi_{2}&=& \frac{1}{\rho_w} \int  \frac{ \ir \omega_s \widehat{p}_{2}(\Kb,f_s) }{- \omega_s^2 +  \ir  g  l} \mathrm{e}^{\mathrm{i}\left[-lz +
    \Theta(\kb,\kpb,s,s')\right]}  {\mathrm d}\Kb   {\mathrm d}f_s 
 \\
p_{2}&=&\int   \frac{\widehat{p}_{2}(\Kb,f_s) }{1 - \ir g  l/\omega_s^2}  \mathrm{e}^{\mathrm{i}\left[-lz +
    \Theta(\kb,\kpb,s,s')\right]} {\mathrm d}\Kb   {\mathrm d}f_s  \label{eq:p2p}
\end{eqnarray}
where $p_{2}$ has been obtained using the linearized version of eq. (\ref{Bernoulli2}). The 
measured pressure signal is the sum of the linear pressure $p_1$, the second-order wave pressure $p_{2}$ given by eq. (\ref{eq:p2p}), and the 
Bernoulli correction $p_{2,B}$ given by
\begin{equation}
 p_{2,B}(z)= \rho_w \sum_{\kb,s,\kpb,s'}  D_{pb} \left(\kb,s,\kpb,s',z\right) Z_{1,\kb}^{s} Z_{1,\kpb}^{s'} 
 \mathrm{e}^{\mathrm{i}
     \Theta(\kb,\kpb,s,s')}\label{P2B}.
\end{equation}
We note that  $p_{2,\mathrm{bot}}$ defined in eq. (\ref{P2bot}) is equal to $p_{2,B}(z=-h)$.

We shall neglect $  g |l|/\omega_s^2$, which is bounded by the ratio between the deep water gravity and sound speeds, which is less than 0.1 
for wave periods less than 180~s. We express
the velocity potential as a sum of propagating (acoustic, $l$ real) and evanescent (acoustic-gravity, $l$ imaginary) modes, 
\begin{equation}
   \phi_2 = \phi_{2,p} + \phi_{2,e}.
 \end{equation}

We get the frequency spectrum of the propagating modes by integrating over 
the inner regions of the wavenumber space (labelled P+S, S and R in figure \ref{fig:4domains}), 
\begin{eqnarray}
F_{p2,p}(f_s)=\int_{K < \omega_s/\alpha_w}  F_{p2,\mathrm{surf}}(\Kb ,f_s) {\mathrm d}\Kb   \label{Fp1Da}.
\end{eqnarray}
For this range of wavenumbers $|k-k'|< K <  \omega_s/\alpha_w$, and using the relations 
 $\omega_s \simeq 4 \pi f$ and, (for small $|f-f'|$), $|k-k'| \simeq 2 \pi |f-f'|/C_g \simeq 8 \pi^2 f |f-f'|$, 
we obtain an upper bound for the frequency difference $ |f-f'| <  g /(2 \pi \alpha_w)$ which is close to= 0.001~Hz. 
Typical ocean wave spectra have a relative frequency half-width $\sigma_f/f$ that is between 0.03 for swells and and 0.07 
for wind-seas \citep{JONSWAP}, 
so that $ E(f) \simeq E(f')$ 
is a good approximation for the interactions that drive long wavelength pressure fluctuations.


The wave spectrum is
thus broad enough for us to evaluate $ F_{p2,\mathrm{surf}}$ at $K=0$ using eq. (\ref{eq:Fp3D}), 
and take it out of the integral in eq. (\ref{Fp1Da}). The acoustic spectrum simplifies to
\begin{equation}
F_{p2,p}(f_s)=\frac{\pi \omega_s^2}{\alpha_w^2} \rho_w^2 g^2  f_s   E^2(f)I(f)   \label{eq:noise_Lloyd}.
\end{equation}
This is identical to the expression given by \cite{Lloyd1981}.

\subsection{Gravity noise in an unbounded ocean}
The pressure associated with acoustic-gravity modes is the other part of the integral in  (\ref{Fp1Da}), 
for $K > \omega_s/\alpha_w$.  The imaginary wave number $l$ gives a   
vertical attenuation of the power spectrum by a factor $ \er^{-2|l|z}$. 
With that attenuation we may, for large enough depths, 
  assume that only modes with $K \ll k$ contribute to the result,  so that we may take 
$F_{p2,\mathrm{surf}}(\Kb,f_s) \simeq F_{p2,\mathrm{surf}}(\Kb=0,f_s)$,  and take it out of the integrand. 
This approximation is valid only up to a maximum wave number $K_{\max}$ that is 
a small fraction of $k$, $K_{\max}=\epsilon k$. For numerical applications we used $\epsilon=0.2$. 

With this approximation we have, 
\begin{eqnarray}
F_{p2,e}(f_s,z)  &= & F_{p2,\mathrm{surf}}(\Kb=0,f_s) 2 \pi \int_{\omega_s/\alpha_w}^{K_{\max}}  K \er^{2|l|z} {\mathrm d} K   \nonumber \\
                 &  = &F_{p2,\mathrm{surf}}(\Kb=0,f_s) 2 \pi \int_{0}^{K_{\max}}  |l| \er^{2|l|z} {\mathrm d} |l|   \nonumber \\
                 &  = &\frac{\pi}{2 z^2}  \rho_w^2 g^2  f_s \left[1-\er^{2 z K_{\max}}\right]  E^2(f)I(f)    \label{Fp1Dag} 
\end{eqnarray}
A previous investigation by \cite{Cox&Jacobs1989} included an extra factor $(1 + z K_{\max})$ in front 
of the exponential term $\er^{2 z K_{\max}}$, because they neglected compressibility effects. 
That term , however, is negligible in the upper part of the water column, and their observations collected within 100 to 
290~m of the surface in 4000~m depth, are thus not affected by this small 
compressibility correction. 

\begin{figure}
\centerline{\includegraphics[width=\columnwidth]{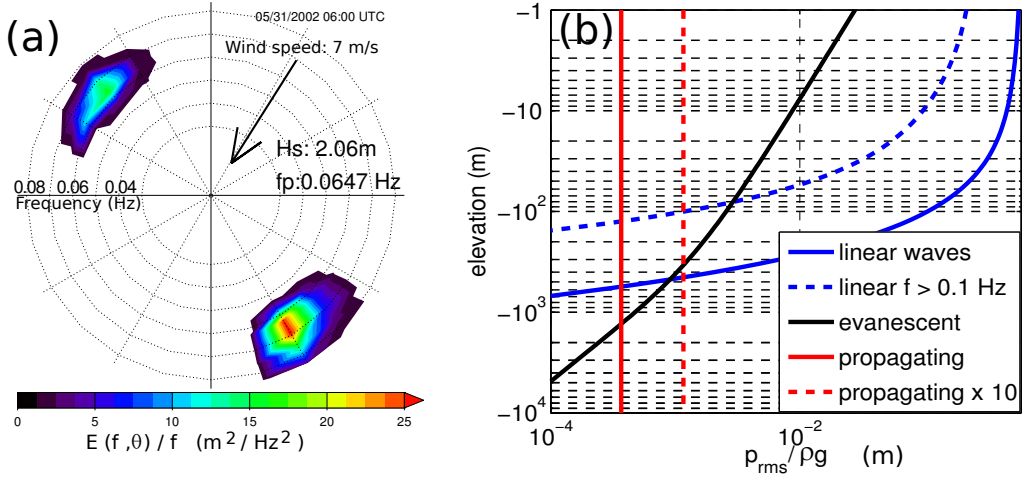}}
  \caption{Example of (a) directional wave spectrum and (b) 
resulting profiles of the different contributions to the pressure fluctuations in the ocean, 
assuming infinite water depth. 
The ratio of double frequency 
to linear wave contributions depends on the amplitude of the waves and on the directional spectral shape, 
because all double 
frequency contributions are proportional to $E^2(f)I(f)$. This directional spectrum was estimated with a numerical wave model, 
and corresponds to the loudest noise 
event recorded at the ocean bottom seismometer H2O, on 31 May  2002 at 25 N, 136 W. This unusual spectrum 
 has large wave energies in opposite directions, radiated from 
a North Pacific storm and Hurricane Alma \citep[This event is analyzed in detail by][]{Obrebski&al.2012}.}
\label{fig:P_profiles}
\end{figure}

As shown in figure \ref{fig:P_profiles}, the oceanic pressure signals can be dominated by linear gravity waves down to 
depths of a few hundred
meters. When looking at the double frequency band, linear waves may only dominate in the top 100~m. 
At these frequencies, the acoustic-gravity modes have the 
most important contribution between about 100 to 500~m, provided that $E^2(f) I(f)$ is large enough. Propagating modes 
should dominate only beyond about 1000 m in the case of an unbounded ocean, or only 300~m, when 
accounting for the reverberation in a finite depth ocean, assuming a typical tenfold amplification for a sea floor with realistic elastic 
properties\footnote{This amplification depends not only on the impedance ratio of the water and crust, which defines 
the amplification coefficients $c_j$ derived below, but also on the seismic attenuation coefficient $Q$, which is discussed in section 4. Realistic calculations following \cite{Ardhuin&al.2011} typically give a
factor 10 to 20 amplification of the sound in the water column due to the bottom elasticity.}. These depths will be reduced in the case 
of surface gravity waves with periods  shorter than the 15~s swells example shown in figure  \ref{fig:P_profiles}.
\subsection{Atmospheric noise source: microbaroms}
The source of noise in the atmosphere can also be derived with the same formalism, as an alternative to the Green functions used by 
\cite{Waxler&Gilbert2006}. Indeed, we may consider the atmospheric motion to be irrotational, so that the equations of motion 
are identical in the atmosphere and in an unbounded ocean, with the only difference that 
 the atmospheric density is $\rho_a$ and the atmospheric sound speed is $\alpha_a$. 
The second-order velocity potential takes the form, 
\begin{equation}
 \phi_{2,a} \propto \exp[\ir(K_{x} x +K_{y} y + l_a z - \omega_s t)] \quad \mathrm{for} \quad z> 0,
\end{equation} 
with
\begin{equation}
 l_a= \sqrt{\frac{\omega_s^2}{\alpha_a^2} - K^2}.
\end{equation}

Because $\rho_w/\rho_a \simeq 1000$, the air motion has only a small $O(\rho_w/\rho_a)$ local 
influence on the water motion, so that the 
solutions derived earlier for the water motion remain valid in the presence of air. 
The air motion, with a velocity potential $\phi_a$ also obeying eq. (\ref{eq:sismo:forcing2lin}) 
is fully determined from the 
water motion via the kinematic boundary conditions on the air and water-sides 
of the interface (\ref{surf_kine_Taylor}), 
\begin{equation}
     \frac{\partial \phi_a }{\partial z} - \frac{\partial \phi }{\partial z} \simeq 
\bnabla \left(\phi_a -\phi\right)\bcdot\bnabla \zeta  - \zeta \frac{\partial^2 \left(\phi_a -\phi\right) }{\partial^2 z} \quad{\mathrm{at}} \quad z=0. \label{surf_kine_Taylora}
\end{equation}
From the first order potential in the air \citep[e.g.][]{Waxler&Gilbert2006}
\begin{equation}
     \phi_{1} = \sum_{\kb,s} \ir s \frac{g}{\sigma}   Z_{1,\kb}^{s}  \er^{-kz} \er^{\ir\left(\kb \bcdot{\mathbf x} - s \sigma\right)} 
\end{equation}
we obtain the second order potential, 
\begin{equation}
      \frac{\partial \phi_{2,a} }{\partial z} =\frac{\partial \phi_2}{\partial z} 
                    + \sum_{\kb,s,\kpb,s'}  D_{za} \left(\kb,s,\kpb,s',z \right) Z_{1,\kb}^{s} Z_{1,\kpb}^{s'} 
\mathrm{e}^{\mathrm{i}
    \Theta(\kb,\kpb,s,s')}\nonumber \\ \label{P2a}
\end{equation}
and a new coupling coefficient
\begin{eqnarray}
D_{za} \left(\kb,s,\kpb,s',z\right)   = -\frac{2 \ir s g}{\sigma} \left(k k' + \kb \bcdot \kpb\right). \nonumber \\ \label{Dza}
\end{eqnarray}

We note that for $\kpb = -\kb$, $D_{za}=0$, so that the long-wavelength motion with $K \ll k$ simplifies to 
\begin{equation}
     \frac{\partial \phi_{2a} }{\partial z} \simeq  \frac{\partial \phi_2 }{\partial z}   \quad{\mathrm{at}} \quad z=0. \label{surf_kine_Tayloratm}
\end{equation}
consistent with the result given by \cite{Posmentier1967} for the interaction of monochromatic wave trains, 
and in disagreement with a factor 8 correction proposed by \cite{Arendt&Fritts2000}. 

This gives a pressure spectrum for the propagating atmospheric waves, 
\begin{equation}
F_{p2,ap}(f_s)=\int_{K < K_{\max}}   \frac{\rho_a^2 \left|l^2\right|}{\rho_w^2 l_a^2 } 
F_{p2,\mathrm{surf}}(\Kb ,f_s) {\mathrm d}\Kb \label{Fp1Dap}  = R(K_{\max}) \frac{\pi \omega_s^2}{\alpha_w^2} \rho_a^2 g^2  f_s   E^2(f)I(f).  \label{eq:noise_barom}
\end{equation}
with the non-dimensional factor
\begin{equation}
R(K_{\max}) = 2 \frac{\alpha_w^2}{\omega_s^2} \int_0^{K_{\max}} \frac{\left|l^2\right|}{l_a^2}   K {\mathrm d}K = = 2 \frac{\alpha_w^2}{\omega_s^2} \int_0^{K_{\max}} \frac{\left|l^2\right|}{l_a^2}   K {\mathrm d}K \label{RKmax}.
\end{equation}
In order to avoid the singularity for $l_a=0$, and atmospheric ducting effects not represented here, we 
take $K_{\max} = \omega_s/( 2 \alpha_a)$ which restricts
the acoustic propagation directions to be within 30 degrees from the vertical. In that case  we have 
$R(K_{\max}) \simeq 0.54$ instead of $R(K_{\max}) =0.25$ with  the vertical propagation approximation of \cite{Waxler&Gilbert2006},
which replaces the $l^2/l_a^2$ factor in the integral by its value $\alpha_a^2/\alpha_w^2$ for $K=0$. Other than that, our expression is consistent with their low Mach number asymptote, 
i.e. $\sigma/k \ll \alpha_a$  \citep[][eq. 61]{Waxler&Gilbert2006}. 
The present theory also allows the estimation of the evanescent wave components given 
by wavenumbers $K > \omega_s/\alpha_a$.


\section{Noise in a finite depth ocean}
For large depths compared to the OSGW wavelength, $kh \gg 1$, the finite depth 
has little effect on the evanescent modes except
for a doubling of the motion amplitude near the bottom, as the vertical profiles of the form 
$\exp(K z)$ are replaced by $\cosh(K z+K h)/\cosh(K h)$. This is similar to the finite depth effect 
on  linear wave motions. 
However, the propagating modes  radiated by the surface will now undergo multiple reflections at the bottom and sea surface, 
as shown in figure \ref{fig:water_and_crust}. The oceanic acoustic field is tightly coupled 
to elastic waves in the crust through these reflections.

One of the greatest complications induced by the presence of a bottom is the heterogeneity of  
the sediment and rock layers below the water column. 
The natural layering of the crust has a strong influence on the sound reflection and 
the nature of the seismic modes \citep[e.g.][]{Latham&Sutton1966,Abramovici1968}. These effects 
will not be considered here, and we follow exactly the theoretical setting of \cite{Hasselmann1963c}.

\subsection{Elastic wave theory}
For simplicity, we assume here that the ocean bottom is a uniform and semi-infinite 
solid, with constant density $\rho_s$ and compression 
and shear wave velocities $\alpha_c$ and $\beta$ given from the Lame coefficients $\lambda$ and $\mu$, 
\begin{subeqnarray}
\gdef\thesubequation{\theequation \textit{a,b}}
  \alpha_c^2 =\frac{\lambda+2 \mu}{\rho_s}, \quad 
  \beta^2 =\frac{\mu}{\rho_s}.
\end{subeqnarray}
Assuming that the crust is a Poisson solid, $\mu=\lambda$ and $\alpha_c = \sqrt{3} \beta$. 

The free wave problem in these conditions was solved by \cite{Stoneley1926}. 
Here we consider the forced problem treated by  \cite{Hasselmann1963c} with a forcing 
by a pressure field $\widehat{p}_{2,\mathrm{surf}}$ at the sea surface, but now generalized to 
an additional wave-induced bottom pressure $\widehat{p}_{2,\mathrm{bot}}$.

The equations of motion in the water column are unchanged from the previous section,
but they are now coupled to the elastic motions of the crust. Crustal
motions can be separated in an irrotational part with a velocity potential $\phi_c$ 
and a rotational part with a stream function $\psi$, both solutions to Laplace's equation. 
With a wave source at the surface, a horizontal propagating wave of phase $\Theta = K x-\omega_s t$, 
implies that $\phi$ and $\psi$ are either decaying or propagating downwards. They must therefore take the following 
form, 
\begin{subeqnarray}
\gdef\thesubequation{\theequation \textit{a,b}}
  \phi_c = A \er^{m(z+h)}\er^{\ir \Theta}, \quad  \psi   = B  \er^{n(z+h)}\er^{\ir \Theta}.\label{eq:phiAB}
\end{subeqnarray}
Both vertical wavenumbers $m$ and $n$ are given by the Fourier transform of eq. (\ref{eq:sismo:forcing2lin}), 
\begin{subeqnarray}
\gdef\thesubequation{\theequation \textit{a,b}}
m =  \sqrt{K^2 - \frac{\omega_s^2}{\alpha_c^2}}, \quad n = \sqrt{K^2 - \frac{\omega_s^2}{\beta^2}} 
\end{subeqnarray}
where the sound speed in water has been replaced by the compression and shear velocities.
For $K > \omega_s/\beta_s$, $m$ and $n$ are real and both compression and shear waves are evanescent. 
For $\omega_s/\alpha_c < K < \beta_s$ the compression wave is evanescent but there is a shear ($S$) wave that propagates through the crust. 

The constants $A$ and $B$ have dimensions of m$^2/$s and are determined by the boundary conditions at the ocean bottom. 

Horizontal and vertical ground displacements are given by the real parts of 
\begin{eqnarray}
\xi_x &=&  \left(K A  \er^{m(z+h)}      + \ir n B  \er^{n(z+h)}\right)  \er^{\ir \Theta} \label{eq:sismo_xi_x}/\omega_s \\
\xi_z &=&    \left(- \ir m A  \er^{m(z+h)} +  K B  \er^{n(z+h)} \right) \er^{\ir \Theta} /\omega_s \label{eq:sismo_xi_z}. \label{eq:xi_z}
\end{eqnarray}

Hooke's law of elasticity gives 
\begin{subeqnarray}
\gdef\thesubequation{\theequation \textit{a,b}}
\tau_{zz}   = \lambda \left(\frac{\partial \xi_x}{\partial x} + \frac{\partial \xi_z}{\partial z}\right) + 2 \mu \frac{\partial \xi_z}{\partial z}, \quad
\tau_{xz}   = \mu  \left(\frac{\partial \xi_x}{\partial z} + \frac{\partial \xi_z}{\partial x}\right)
\end{subeqnarray}
The zero tangential stress on the ocean bottom $\tau_{xz}(z=-h)=0$ yields 
the following relationship between $A$ and $B$, which is typical of Rayleigh waves, 
\begin{eqnarray}
 B= \frac{2 \ir K m}{n^2+K^2} A.
\end{eqnarray}

Thus, in addition to the unknown water-side amplitudes $C$ and $D$ of the velocity potential at the bottom, we have one more unknown, the compression wave amplitude $A$ on the solid side. 

The three equations that relate $C$, $D$ and $A$ are:  the combined kinematic and dynamic boundary condition (\ref{eq:eq_phi_2}), 
and the bottom continuity of normal velocity 
\begin{equation}
 \frac{\partial \phi_2}{\partial z} = \frac{\partial \phi_c}{\partial z}  +\frac{\partial \psi}{\partial x} \quad {\mathrm{at}} \quad z=-h  
\end{equation}
and normal stress, 
\begin{equation}
 -\tau_{zz}(-h)= p(-h) = \rho_w \frac{\partial \phi_2(-h)}{\partial t} + \widehat{p}_{2,\mathrm{bot}}.
\end{equation}
For waves in intermediate or shallow water, i.e. $kh < \pi$, the Bernoulli term $\widehat{p}_{2,\mathrm{bot}}$ that was 
not considered by \cite{Hasselmann1963c} should be included. We thus obtain the linear system of equations
\begin{eqnarray}
   \left( - \omega_s^2 {\rho_w} +g \ir l\right) \er^{\ir lh}   C         & +  \left( - \omega_s^2 {\rho_w} -g \ir l\right) \er^{-\ir lh}     D       & =  \frac{\ir \omega_s}{\rho_w}\widehat{p}_{2,\mathrm{surf}}(\Kb,f_s) \\
 q A   - \ir l C               & +  \ir l D               & =  0 \\
 r A  - \ir \omega_s  {\rho_w} C& -  \ir \omega_s {\rho_w} D & =  -  \widehat{p}_{2,\mathrm{bot}}(\Kb,f_s)
\end{eqnarray}
with   
\begin{eqnarray}
q&=&\frac{m \omega_s^2}{\omega_s^2-2K^2 \beta^2} \\
r&=& \frac{\ir}{\omega_s} \rho_s\left[- \frac{4 \beta^4 K^2 m n}{\omega_s^2- 2K^2 \beta^2}
 + \left(\omega_s^2- 2K^2 \beta^2 \right)\right] \nonumber \\
 &=&  \frac{\ir}{\omega_s} \left[ - \rho_s  m^2 \alpha_c^2  
             + \lambda  k^2 + 4 \mu \frac{K^2 m n}{n^2+K^2} \right] \label{eq:syst3}
\end{eqnarray}
Since we are in the range where $l<\omega_s /\alpha_w$ we may neglect again $g |l|/\omega_s^2$, which is less than 0.1 
for OSGW periods less than 180~s.
We rewrite these equations in matrix form,
\begin{equation}
M  [ A, \quad C, \quad D]^T =  [- \widehat{p}_{2,\mathrm{surf}}(\Kb,f_s) , \quad  0, - \widehat{p}_{2,\mathrm{bot}}(\Kb,f_s)]^T \label{eq:matrix}
\end{equation}
with 
\begin{equation}
 M =\left[\begin{array}{ccc}
 0 &\  \ir \rho_w \omega_s \er^{\ir lh}            & \ir \rho_w \omega_s  \er^{-\ir lh}  \\  
 q &- \ir l                   &  \ir l \\
 r &  - \ir \omega_s  {\rho_w}  &  - \ir \omega_s  {\rho_w}  \\
 \end{array}\right]
\end{equation} 

The general solution of eq. (\ref{eq:matrix}) is the sum of one particular forced solution and 
the general solution of the homogeneous system, without the right hand side forcing, i.e. the 
free waves. This combination of free and forced waves is fully determined by the initial conditions.

The forced solution is readily expressed using the determinant of the system 
\begin{eqnarray}
\det(M) &=& - \rho_w \omega_s \left(l \er^{\ir lh} r   + \omega_s  {\rho_w} \er^{-\ir lh} q   
                           - l  \er^{-\ir lh} r    -  \omega_s  {\rho_w} \er^{\ir lh} q\right) \\
 &=& \frac{2 \rho_w }{ \left (\omega_s^2- 2  K^2 \beta^2 \right)}\left\{ l \rho_s \cos(lh) \left[4 \beta^4 K^2 m n - \left (\omega_s^2- 2  K^2 \beta^2 \right)^2  \right]
   - \rho_w m \sin(lh) \omega_s^4 \right\},\nonumber \\
\end{eqnarray}
in the form
\begin{subeqnarray}
  \gdef\thesubequation{\theequation \textit{a}}
  A&=& 2 l \rho_w  \omega_s \frac{\widehat{p}_{2,\mathrm{surf}}(\Kb,f_s) +  \cos(lh) \widehat{p}_{2,\mathrm{surf}}(\Kb,f_s)}{ \det (M)} \label{PtoA}\\
  \gdef\thesubequation{\theequation \textit{b}}
  C&=&  \frac{- \ir \left( l r + q  \omega_s \rho_w\right) \widehat{p}_{2,\mathrm{surf}}(\Kb,f_s) + q \er^{-\ir lh}  \widehat{p}_{2,\mathrm{bot}}(\Kb,f_s)}{\det(M)} \\
  \gdef\thesubequation{\theequation \textit{c}}
  D&=& \ir \left( q  \omega_s \rho_w - l r  \right) \frac{\left( q  \omega_s \rho_w - l r  \right)  \widehat{p}_{2,\mathrm{surf}}(\Kb,f_s) 
-  q \er^{-\ir lh}   \widehat{p}_{2,\mathrm{bot}}(\Kb,f_s) }{ \det(M)}.
\end{subeqnarray}
As detailed below, this determinant vanishes for the pairs $(\omega_s,\Kb)$ that fall on the 
dispersion relation of Rayleigh waves. But the solution for random waves can always be obtained by 
integrating across this singularity, following  \cite{Hasselmann1962}. 

For $kh \gg 1$ we may neglect $\widehat{p}_{2,\mathrm{bot}}$ and, 
\begin{subeqnarray}
\gdef\thesubequation{\theequation \textit{a}}
C&=& \frac{\widehat{p}_{2,\mathrm{surf}}(\Kb,f_s)}{\rho_w \omega_s}\frac{\ir r' - q'}{2r' \cos(lh) - 2q'\sin(lh)} \\
\gdef\thesubequation{\theequation \textit{b}}
D&=& \frac{\widehat{p}_{2,\mathrm{surf}}(\Kb,f_s)}{\rho_w \omega_s}\frac{\ir r' + q'}{2r' \cos(lh) - 2q'\sin(lh)} 
\end{subeqnarray}

with 
\begin{subeqnarray}
\gdef\thesubequation{\theequation \textit{a,b}}
q'= \rho_w \omega_s q, \quad  r'=  \ir l r.  
\end{subeqnarray}

\subsection{Acoustic noise in a bounded ocean}
Taking typical values of the water and crust density and sound speeds gives  $r' / q' > 70$ for the free modes that are significantly generated by 
the waves (i.e. $c_j > 0.1$, as defined below by eq. \ref{eq:c_j}). We can thus consider that $q'/r' \ll 1$, 
which gives $C \simeq D \simeq \widehat{p}_{2,\mathrm{surf}}(\Kb,f_s)/[2 \rho_w \omega_s (\cos (lh)+q'/r')]$.
The velocity potential and pressure in the water are given by,
\begin{subeqnarray}
\gdef\thesubequation{\theequation \textit{a,b}}
 \phi_2 \simeq  \ir  \widehat{p}_{2,\mathrm{surf}}(\Kb,f_s) \frac{ \cos[l(z+h)-q'/r']}{ \rho_w \omega_s (\cos (lh)-q'/r')} \er^{\ir \Theta}, \quad
 p_2 \simeq   \ir \widehat{p}_{2,\mathrm{surf}}(\Kb,f_s) \frac{ \cos[l(z+h)] -q'/r'}{ (\cos (lh)-q'/r')} \er^{\ir \Theta}.\nonumber \\
\end{subeqnarray}
The small but finite factor $q'/r'$ ensures that the solution remains finite as a small fraction of the acoustic energy 
is radiated into the crust, otherwise the acoustic energy would accumulate in the water column. The pressure 
oscillations are thus maximum for resonant frequencies such that the ratio of the water depth and 
vertical wavelength $l h/(2 \pi)$ is $1/4$, $3/4$, $5/4$ ... 
 We note that the vertical wavelength $2 \pi/l$ is always greater than the acoustic wavelength 
$2 \pi/\sqrt{K^2+l^2}= f_s \alpha$. As a result, the resonant frequencies are shifted to higher values
 compared to the vertical resonant condition that is given
by  $f_s h  \alpha_w = 1/4$ ... 

We may now integrate the pressure spectrum for all acoustic wavenumbers to find, again, the 
frequency spectrum, 
\begin{equation}
F_{p2,ap}(f_s)=2 \pi \rho_w^2 g^2  f_s   E^2(f)I(f)  \int_0^{\omega_s/\alpha_w} \left[\frac{\cos(lz+lh) -q'/r'}{\cos (lh)-q'/r'}\right]^2 K {\mathrm d} K.\nonumber \\
\label{eq:sound_finite}
\end{equation}
We now have a depth dependence of the sound spectrum. At the surface it is equal to the widely used unbounded ocean value \citep{Lloyd1981},
but in the water column it can be strongly amplified at depths where $\cos(lh)$ approaches 0,
 which includes the near-bottom region where all the resonant modes have an anti-node.

\subsection{Rayleigh waves}
 In order to simplify the algebra, we consider in this section waves in deep water and neglect bottom forcing.  
As a result, this section brings no new results compared to \cite{Hasselmann1963c}, but the properties are discussed in more detail and more explicit expressions are presented 
that will be used later in the derivation of new solutions for other types of seismic waves.

\begin{figure}
\centerline{\includegraphics[width=\columnwidth]{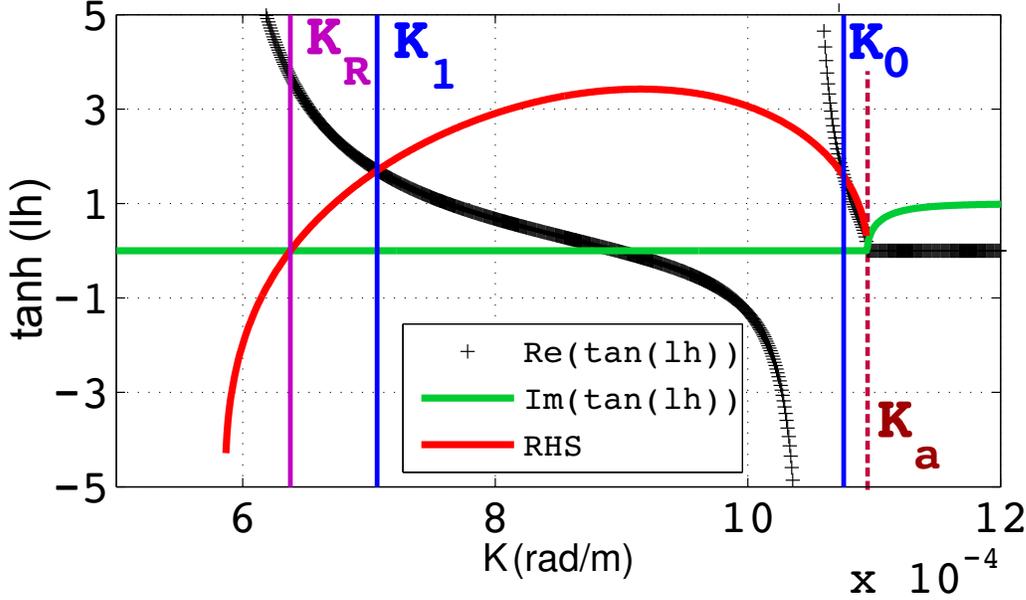}}
  \caption{Wavenumbers of seismic and acoustic modes for a fixed frequency, 
which are the solutions of eq. (\ref{Rayleigh_dispersion}). The right hand side of 
that equation is abbreviated as RHS. 
Here we used $h=5000$~m, $\alpha_w=1500$ m/s, $\beta=3200$m/s, $\rho_w=1000$~kg/m$^3$ and $\rho_s=2500$~kg/m$^3$. 
This graph corresponds to 
a seismic frequency $f_s=0.263$~Hz, for which $f_s h/\alpha_w=0.88$. 
Here the determinant has three real roots, one acoustic mode $K_a$ for which $l=0$ (horizontal propagation)
and two seismic modes $K_0$ and $K_1$. 
The number of these roots increases with the frequency. One new solution appears every time $2\pi f_s$ 
becomes larger than $\omega_{s,j}^0$ 
 defined by eq. (\ref{omega_cj}).
For reference we also give the wavenumber for which the right hand side of eq. (\ref{Rayleigh_dispersion}) 
is zero, which is the Rayleigh wavenumber $K_R$ in the absence of 
the ocean layer ($h=0$).}
\label{fig:Rayleigh_roots}
\end{figure}

For a fixed frequency $\omega_s$, there is at least one  
wavenumber $K$ for which $\det(M)=0$. This condition defines the dispersion relation of the Rayleigh modes \citep{Stoneley1926}, 
\begin{equation}
 \tan  \left(  l h \right)=    \frac{l \rho_s}{m \rho_w} 
\times \frac{4 \beta^4 K^2 m n  - \left (\omega_s^2- 2  K^2 \beta^2 \right)^2   }{ \omega_s^4 } \label{Rayleigh_dispersion}
\end{equation}
with the fundamental mode corresponding to the largest $K$ value.
Figure \ref{fig:Rayleigh_roots} illustrates the family of modes for a give frequency. 

\begin{figure}
\centerline{\includegraphics[width=0.7\columnwidth]{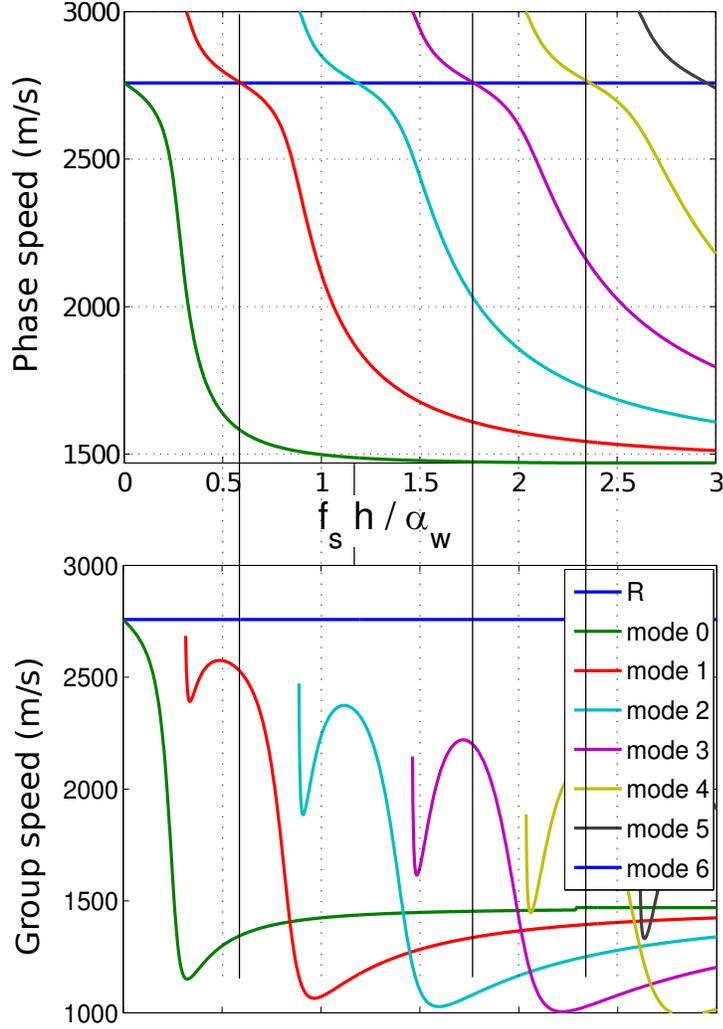}}
  \caption{Phase (top) and group (bottom) speeds of
Rayleigh waves as a function of the dimensionless water depth. Calculations used $\alpha_w=1470$~m~s$^{-1}$, $\beta=3000$m~s$^{-1}$, 
$\rho_w=1000$ and $\rho_s=2500$~kg~m$^{-3}$. Values without the water layer are also indicated with the blue line. In that case 
Rayleigh waves are not dispersive.  }
\label{fig:Rayleigh_dispersion}
\end{figure}

The phase speeds of the Rayleigh modes vary continuously from the shear wave velocity $\beta$, in the limit 
$n=0$ where the shear waves transition from propagating to evanescent, to the sound speed 
in water $\alpha_w$, in the limit $l=0$ where the acoustic modes become evanescent in the water (figure \ref{fig:Rayleigh_dispersion}.a). This variation of the phase speed
has an inflexion point close to the phase speed of Rayleigh waves without the water layer, corresponding to a maximum in group speed (figure \ref{fig:Rayleigh_dispersion}.b).  

We may now use eq. (\ref{PtoA}) to obtain the 
ground displacement amplitude 
\begin{equation}\delta \equiv \xi_z(-h) = G(K_x,K_y,\omega_s) \widehat{p}_{2,\mathrm{surf}}(K_x,K_y,\omega_s)\end{equation}
as a function of  the amplitude of the sea surface pressure, with the 
transfer function
\begin{equation}
G = \frac{2 \ir \rho_w l  m  (K^2-n^2 )}{(n^2+K^2) \det (M)}.
\end{equation}

\begin{figure}
\centerline{\includegraphics[width=0.8\columnwidth]{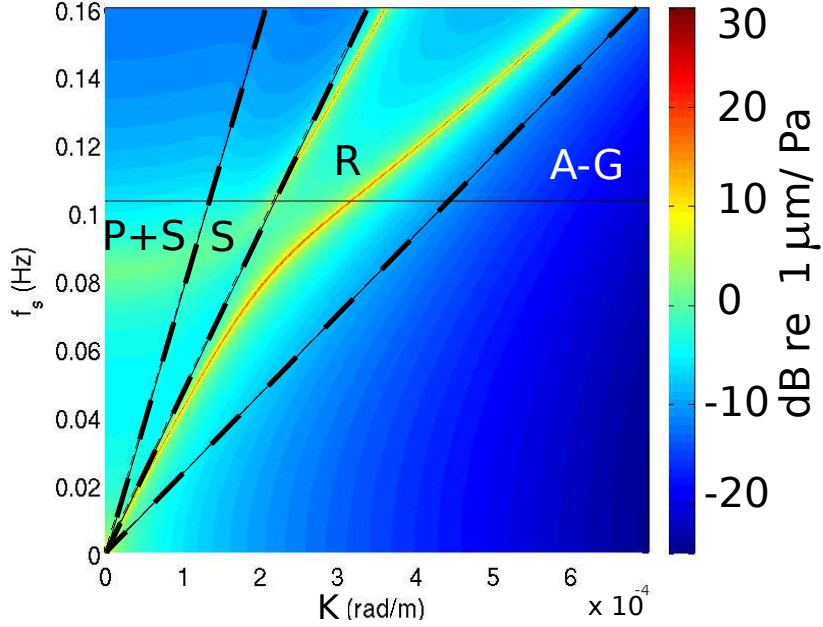}}
  \caption{Magnitude of the transfer function $G(K,f_s)$ illustrating the 
singularities along the dispersion relations of the free Rayleigh modes with wavenumbers $K=K_{n}(f_s)$, 
that correspond to $\det(M)=0$. The oblique 
dashed lines correspond to phase speeds equal to $\alpha_c$, $\beta$, and $\alpha_w$, and separate the four domains 
body waves (P+S), mixed body and evanescent waves (S), Rayleigh waves (R) and acoustic-gravity modes (A-G).
}
\label{fig:k_f_plot}
\end{figure}

An example of this transfer function is shown in figure \ref{fig:k_f_plot}, in which the dispersion relation of
the Rayleigh modes appear as the narrow red bands around the singularities of $G$. 

To evaluate the complete family of Rayleigh wave solutions with dispersion relations 
$\omega_{s,j}(K)$, where $j$ is the mode number, we need to examine the nature of these singularities.  We re-write 
the determinant of the system as, 
\begin{equation}
\det(M) =  \frac{2 \ir l \rho_s \rho_w \cos(lh)}{ \left (\omega_s^2- 2  K^2 \beta^2 \right)}  
       \left\{ \left(4 \beta^2 k^2\right)\left(m n +\omega_s^2 
- \beta^2 K^2\right)  - \omega_s^4  \left(1+\tan(lh) \rho_w m/{\rho_s l} \right) \right\}.
\end{equation}

For most frequencies, the singularities are simple, allowing a Taylor expansion of $G$ in this form 
\begin{equation}G(K,\omega_s) \simeq \frac{G'(K_j(\omega_s))}{\omega_s^2-\omega_{s,j}^2(K)}.\end{equation}

However, for each mode $j$, there is a critical frequency
\begin{equation}
 \omega_{s,j}^0=\frac{1}{h}\left[\arctan(-\frac{l}{m}{\rho_s}{\rho_w}) + j \pi \right] \label{omega_cj}
\end{equation}
for which both  $\tan(lh) {\rho_w m}/{\rho_s l}  = 1$,  
and $\omega_s^2- K^2 \beta^2=0$. These critical frequencies $\omega_{s,j}^0$ exist for modes $j>0$. 
These frequencies are those for which  new Rayleigh modes 
appear, in a way similar to the Love waves discussed by Aki \& Richards (2002, figure 7.3)\nocite{Aki&Richards2002}. 
The singularities at $(K,\omega_s)=(\omega_{s,j}^0/\beta,\omega_{s,j}^0)$ are not
simple and for these we have $G(\omega_s,k) \simeq G'(K_j(\omega_s))/\sqrt{K^2-K_j^2(\omega_s)}$.  

Solutions for the vertical  displacement at the top of the crust have a spectral amplitude 
which is linearly related to the equivalent surface pressure amplitude,  
\begin{equation}
\delta(K_x,K_y,\omega_s) = G(K_x,K_y,\omega_s) \widehat{p}_{2,\mathrm{surf}}(K_x,K_y,\omega_s).\label{PtoDelta}
\end{equation}

For the simple singularities of $G$, 
we may write  $G(K,\omega_s) = G'(K,\omega_{s,j})/(\omega_s^2-\omega_{s,j}^2) + O(\omega_s- \omega_{s,j})$. 
Taking initial conditions 
$\delta(t=0)=0$ and $\partial \delta / \partial t=0$ gives the full solution \citep[][eq. 3.2]{Hasselmann1962} 
of the $j^{\mathrm th}$ Rayleigh mode response
\begin{equation}
\delta(K_x,K_y,\omega_s) = G'(K,\omega_{s,j})\er^{\ir K x} 
 \left[\frac{1}{\omega_{s,j}^2-\omega_s^2} 
\er^{-\ir \omega_s t}-\frac{1}{2 \omega_{s,j}}\left(\frac{\er^{\ir \omega_{s,j} t}}{\omega_{s,j} - \omega_s} 
+ \frac{\er^{\ir \omega_{s,j} t}}{\omega_{s,j} + \omega_s} \right) \right],
\end{equation}
as a function of $K$, where $\omega_{s,j}(K)$ is frequency of  the $j^{\mathrm{th}}$ Rayleigh mode. 
For a forcing that varies slowly on the scale of the seismic period $2 \pi / \omega_s$, and provided that the forcing 
spectrum is continuous in spectral space, this gives the rate of change of the ground displacement spectrum given by \cite{Hasselmann1963c}, 
\begin{equation}
\frac{\partial F_\delta(K_x,K_y)}{\partial t} =  S_{DF}(K_x,K_y)=\frac{\pi \left|G'\right|^2}{2 \omega_{s,j}^2}F_{p2,\mathrm{surf}}(K_x,K_y,\omega_s).
\label{SDF_k}
\end{equation}

The other discrete singularities,  at $(K,\omega_s)=(\omega_{s,j}^0/\beta,\omega_{s,j}^0)$,  are associated with 'conical' or 
'head' waves  \citep{Aki&Richards2002}, for which the vertical 
wavenumber $n=0$, and that propagate along the ocean-crust 
interface. These horizontally propagating shear waves  are generated with an evanescent compression wave. 
The singularity is integrable over the two spectral dimensions $\omega_s$ and $K$.

Although it looks like only the resonant forcing contributes to the solution, it is in fact the near-resonant forcing 
($\omega_s \simeq \omega_{s,j}$)
that builds up the seismic noise, because the exact resonant terms have a zero measure in spectral space. Indeed, a purely resonant 
forcing would give an amplitude that increases linearly with time, and an energy that increases like $t^2$. 
The linear growth of energy in time can be interpreted as an effect of the narrowing with time of the frequency bandwidth in which the 
interaction is significant. 
This is a general property of wave-wave interactions \citep[see also][]{Hasselmann1966}.

\begin{figure}
\centerline{\includegraphics[width=0.7\linewidth]{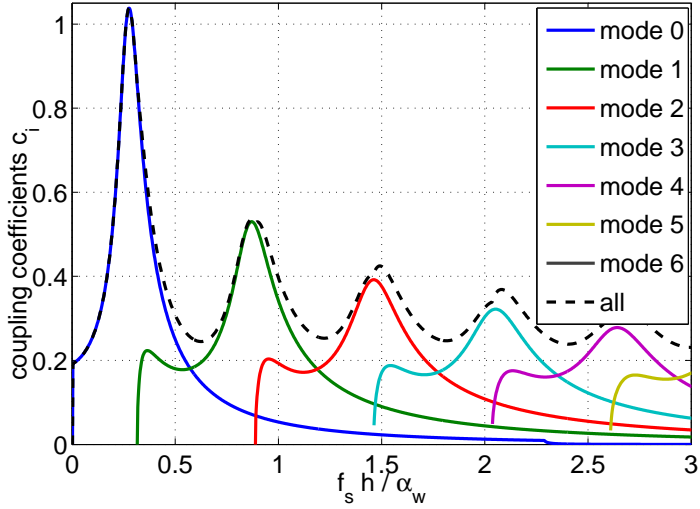}}
  \caption{Dimensionless coefficients  $c_j$ that amplify the wave-induced pressure into ground displacement. 
  The maxima of $c_j$ correspond to quarter-wavelength resonance typical of organ pipes, except that here 
the sound waves propagate obliquely in the water column, which is why the maxima are at  values of $f_s h/ \alpha_w$
which are not exactly at 1/4, 3/4..., but shifted to higher frequencies by a factor $l/\sqrt(K^2+l^2)$. 
The amplitudes of the peaks depend on the impedance ratio of the sea water and crust. Hence the 
peak amplitude increases with $\rho_s \beta/(\rho_w \alpha_w)$. For example,   $\beta=2800$~m~s$^{-1}$, gives a maximum of  0.88 
for $c_0$
instead of 1.03 here with $\beta=3000$~m~s$^{-1}$.}
\label{fig:sismo_coef}
\end{figure}

We express the source of seismic noise with the rate of increase of the variance of $\delta$ per unit of propagation distance. It is 
\begin{equation}
 S_{DF}(\omega_s)=\frac{K(s) S_{DF}(K_x,K_y)}{U^2} =\frac{ 4 \pi^2 f_s c^2_j}{\beta^5 \rho_s^2} F_{p2,\mathrm{surf}}(K_x,K_y,\omega_s)
\end{equation}
where $U$ is the group speed of the seismic waves, 
and $c_j$ is a dimensionless coefficient that depends on $\omega_s h/\alpha_w$ and the seismic mode index $j$, 
shown in figure \ref{fig:sismo_coef},
\begin{equation}
c_j^2  =  \frac{\beta^5 \rho_s^2 K_j}{ U_j^2 2 \pi \omega_s } \frac{\pi \left|G_j'\right|^2}{2 \omega_s^2}.\label{eq:c_j}
\end{equation}
We note that a missing $2 \pi$ in eq. (5) of \cite{Ardhuin&al.2011} has been corrected here. 

A very rough simplification can be obtained by taking $U$ and $K$ independent of $j$ and $h$. 
Then $S_{DF}(f_s)$ can be taken as a sum of all Rayleigh modes in the form 
\begin{eqnarray}
S_{DF}(f_s) =  \frac{4 \pi^2 f_s}{\beta^5 \rho_s^2} \left(\sum_{i=0}^{\infty} c_j^2 \right)F_{p}(\mathbf{K} \simeq 0 ,f_s).
\label{eq:sismo_c_n} \nonumber \\
\end{eqnarray}
Values of $c_j$ are obtained from eq. (\ref{eq:c_j}), and shown in figure \ref{fig:sismo_coef}.

We may propagate these sources of seismic waves in a vertically symmetric earth model, neglecting  all three-dimensional 
propagation effects, and parameterizing seismic wave scattering and dissipation 
 with a uniform quality factor $Q$. Under these assumptions, the spectral density of the vertical ground displacement at
$z=-h$ and at the longitude $\lambda_O$ and latitude $\phi_O$
\begin{equation}
F_\delta(\lambda_O,\phi_O,f_s)=  \int_{- \pi/2}^{ \pi/2} \int_0^{2 \pi}  \frac{ S_{DF}(f_s)}{R_E \sin \Delta} ~ \mathrm{e}^{-2 \pi  f_s \Delta R_E / (U Q)} 
 (R_E^2 \sin \phi_O' {\mathrm d} \lambda_O' 
{\mathrm d} \phi_O')  \label{F_delta}
\end{equation}
with $R_E$ the Earth radius, and $U$ the seismic group velocity. 
The term $(R_E^2 \sin \phi_O' {\mathrm d} \lambda_O' 
{\mathrm d} \phi_O')$ is the Earth surface area element. The denominator  $(R_E \sin \Delta)$ 
is the geometrical spreading factor for wave energy that follows geodesics on the sphere \citep[e.g.][]{Kanamori&Given1981}, replacing the 
distance $(R_E \Delta)$ used in flat Earth models \citep[e.g.][]{Hasselmann1963c}. 

The Rayleigh waves thus generated propagate away like free modes, those that exist without the local forcing. For these free modes 
with a monochromatic ground displacement $\delta(x,y,t)$
the surface pressure is constant so that $C=-D$ and the velocity potential and pressure in the water take the form, 
\begin{subeqnarray}
\phi_2(z)&=&  \frac{\omega_s}{l \cos(lh)} \sin(lz)  \delta(x,y,t)\\
p_2(z)&=&  \ir \rho_w \frac{\omega_s^2}{l \cos(lh)} \sin(lz)  \delta(x,y,t) \label{p_free_Rayleigh}.
\end{subeqnarray}

For varying water depths, and on land, one may assume that the seismic energy is propagated along a ray and apply the 
refraction coefficient given by \cite{Hasselmann1963c}.

\subsection{Seismic $P$ and $S$ waves}
Unlike the Rayleigh waves that grow resonantly in time owing to the trapping in the ocean/crust waveguide, the $P$ and $S$ waves, 
for which the vertical wavenumbers $m$ or $n$ are complex, 
radiate into the earth's interior, and their energy level is given directly by its value at the source. 
In particular, for $ K < \omega_s/\alpha_c$ we have propagating $P$ waves with a velocity potential amplitude $A$ 
given by eq. (\ref{PtoA}).  In the particular case $K=0$, which correspond to standing waves, we have only have $P$ waves, 
no $S$ waves or Rayleigh waves, and these propagate exactly along the vertical axis. 
The frequency spectrum of vertical ground displacements at $(z=-h)$ can be evaluated directly with (\ref{PtoDelta}) because 
$G$ has no singularity in this range of wavenumbers, 
\begin{equation}
F_{\delta,P}(f_s)= f_s   E^2(f)I(f)  \frac{\rho_w^2 g^2}{\rho_s^2 \beta_s^4 } c_P^2 
\label{eq:P_source}
\end{equation}
with a non-dimensional coefficient $c_{P}$,
\begin{equation}
 c_P^2=2 \pi \int_0^{\omega_s/\alpha_c} 
\frac{4 l^2 m^2 \rho_s^2 \beta_s^4}{\omega_s^2 \det^2 (M)} K {\mathrm d} K \label{c_P_def}.
\end{equation}

\begin{figure}
\centerline{\includegraphics[width=0.7\linewidth]{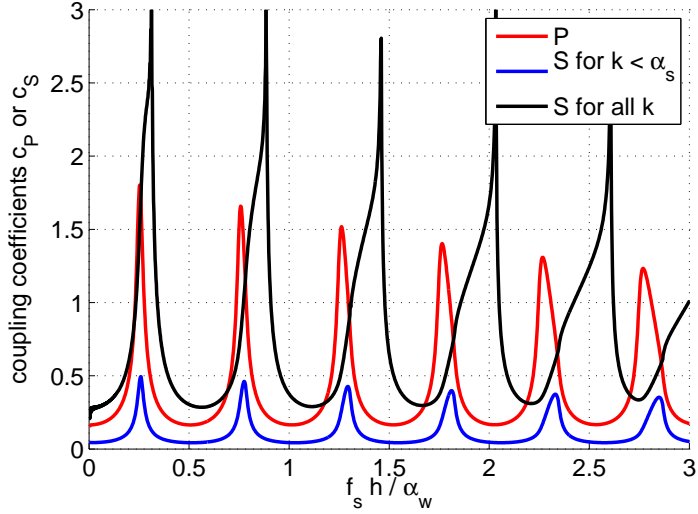}}
  \caption{Non-dimensionless coefficients  $c_P$ and $c_S$ that amplify the wave-induced pressure into ground displacement
associated with $P$ and $S$ waves.}
\label{fig:sismo_coefPS}
\end{figure}
A similar expression can be written for $S$ waves, and both are illustrated in 
figure  \ref{fig:sismo_coefPS}, 
\begin{equation}
F_{\delta,S}(f_s)= f_s   E^2(f)I(f)   \frac{\rho_w^2 g^2  }{\rho_s^2 \beta_s^4 } c_S^2.
\label{eq:S_source}
\end{equation}

 However, in the range of wavenumbers 
where $S$ waves exist, $k < \omega_s/\beta$, there can also be evanescent $P$ waves, and the system can approach the 
singularity for $\omega_s=\omega_{s,j}$ and $k=\omega_s/\beta$. We evaluated numerically the coefficient 
\begin{equation}
 c_S^2=2 \pi \int_0^{\omega_s/\beta} 
\frac{4 l^2 m^2 k^2 \rho_s^2 \beta_s^4}{\omega_s^2 (n^2+k^2)\det^2 (M)} K {\mathrm d} K.
\end{equation}
Due to the typically three times stronger attenuation of $S$ waves compared to $P$ waves in the Earth mantle \citep[e.g.][]{Anderson&Hart1978,Pasyanos&al.2009}, we will now focus 
on $P$ waves only, which should dominate in the far field of the noise source. 
\begin{figure}
\centerline{\includegraphics[width=0.65\linewidth]{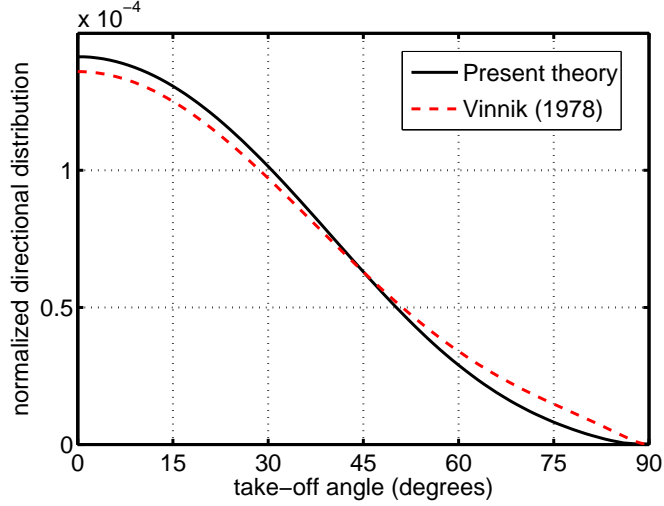}}
  \caption{Dimensionless coefficients  $c_{P,\varphi}$ that amplify the wave-induced pressure into ground displacement. The 
maximum for a zero take-off angle corresponds to vertically propagating compression waves, and the compression waves that propagate 
along the crust have a vanishing amplitude.}
\label{fig:sismo_coef_CPphi}
\end{figure}

For the estimation of the spectrum recorded outside of a source area, it is more convenient to 
express the local seismic source as a function of the horizontal propagation angle $\theta$, and the vertical take-off angle 
$\varphi$. For $P$ waves, this gives, 
\begin{equation}
F_{\delta,P}(f_s,\theta,\varphi)= f_s   E^2(f)I(f)  \frac{1}{\rho_s^2 \beta_s^4 } c_{P,\varphi}^2 \sin \varphi
\label{eq:P_wave_at_source}
\end{equation}
with the non-dimensional coefficient $c_{P,\varphi}$ defined by 
\begin{equation}
c_{P,\varphi}^2=
\frac{4 l^2 m^2  \rho_s^2 \beta_s^4}{\omega_s \alpha_c  \det^2 (M)}\frac{\partial K}{\partial \varphi}, 
\end{equation}
which is the normalized source per unit solid angle $\Omega$, so that the average over the half space of downward directions 
${\Omega^-}$ is
\begin{equation}
c_P^2 = \int_0^{2 \pi} \int_0^{\pi/2} c_{P,\varphi}^2 \sin \varphi {\mathrm d} \varphi  {\mathrm d}\theta  =  \int_{\Omega^-} c_{P,\varphi}^2  {\mathrm d} \Omega
\end{equation}
as defined by eq. (\ref{c_P_def}). 

It is noteworthy that the distribution of the $P$-wave energy with the take-off angle is very  close to the 
one given by a small disk pushing at the top of a uniform half space, as given by \cite{Miller&Pursey1955} and used by 
\cite{Vinnik1973}, although it also varies with the non-dimensional water depth $f_s h/\alpha_w$. 
The only missing item in the work by \cite{Vinnik1973} is the very strong amplification of the motion 
for resonant frequencies associated with the water layer. Due to the large impedance contrast at the water-crust interface 
the relative amplification of $P$ waves is one order of magnitude stronger 
than for Rayleigh waves. We thus expect a much tighter correspondence of the strong seismic noise sources with 
the water depths that correspond to a maximum amplification.

\subsection{Observable $P$ wave spectra}
We will now finish our analysis of these body waves by estimating the ground motion due to $P$ and Rayleigh waves as a function 
of the distance from the source, an application of practical interest.  
 For a seismic station or hydrophone in the ocean, the incoming energy at the receiver will 
 amplified by multiple bottom and surface reflections in the water column. However, this $P$-wave signal 
from remote sources is likely to be dwarfed by locally generated noise. 

For a land-based station, we may assume that $P$ waves arrive directly from the source area. There are many other 
seismic wave phases that have undergone multiple reflections at the surface, these are called $PP$, $PPP$ ..., or at any inner 
boundary of the Earth, like $PkP$ phases that have reflected off the mantle-core interface \citep{Aki&Richards2002}. These can be treated exactly like the 
direct $P$ phase. 
 
On arrival at the receiver, these waves are totally reflected at the Earth surface at $z=0$, 
which doubles the ground motion, 
so that the ground displacement is given by the integral of four times the incoming spectral densities over the 
directions $\theta$ and $\varphi$, which can be replaced 
by an integral over the source positions $(\lambda_S,\phi_S)$. The transformation from the ray parameters $(\theta,\varphi)$ 
to the geographical coordinate can be obtained approximately for any type of seismic wave using travel time tables
\citep[e.g.][]{Snoke2009}, which also provide the travel time $\tau$. 

\begin{figure}
\centerline{\includegraphics[width=0.7\linewidth]{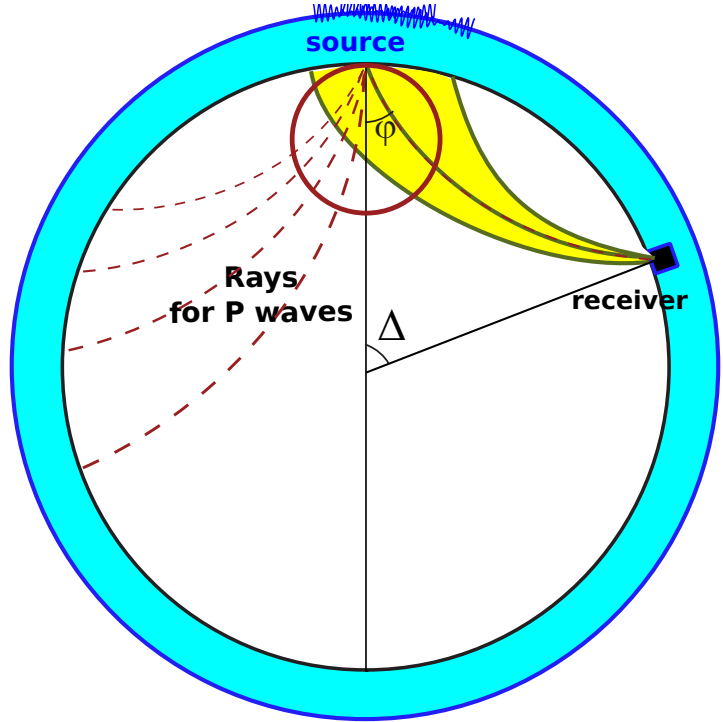}}
  \caption{Schematic of rays for seismic $P$ waves radiated from a point source (red dashed lines) with a directional 
distribution $c_{P,\varphi}$ (solid red line), and ensemble of rays received at a given station from an extended noise source
 (yellow half-banana).}
\label{fig:P_wave_propagation}
\end{figure}

We may now express the ground displacement due to $P$ waves at the observing station of coordinates $(\lambda_O,\phi_O)$, 
as a function of the same quantity at the location of sources, as given by eq. (\ref{eq:P_wave_at_source})
\begin{equation}
F_{\delta,P}(\lambda_O,\phi_O,f_s) =\int_{\Omega^+} 4 F_{\delta,P}(\lambda,\phi,f_s,\theta,\varphi) 
 \mathrm{e}^{-2 \pi  f_s \tau(\lambda_S,\phi_S) / Q} {\mathrm d} \Omega ,   \label{F_delta_P}
\end{equation}
with ${\mathrm d} \Omega$ an element of solid angle that corresponds to the ensemble of rays arriving 
from an Earth surface element around the sources located at $(\lambda_S,\phi_S)$. This sum may also be transformed 
as an integral over the ocean surface by properly mapping $\Omega$ to $(\lambda_S,\phi_S)$. 
The elementary solid angle ${\mathrm d} \Omega$  is zero for the so-called 
shadow zones, the regions for which there is no $P$-wave ray that connects to the observing station. 
For a single phase of seismic waves, this ensemble of rays has a half-banana shape, as illustrated on figure \ref{fig:P_wave_propagation}.

\begin{figure}
\centerline{\includegraphics[width=0.7\linewidth]{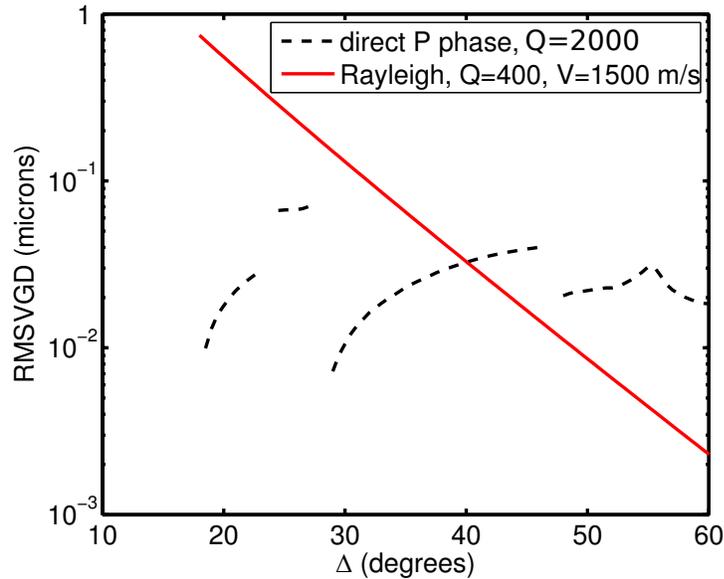}}
  \caption{Estimates of the rms vertical ground displacement associated with Rayleigh or $P$ waves, 
as a function of the epicentric angle $\Delta$, for a source of intensity 
$\int F_{p2,\mathrm{surf}}(\Kb = 0,f) \mathrm{d}f = 4.2\times 10^4$ 
hPa$^2$~m$^2$ over a 330 by 330 km square, 
assuming an attenuation factor $Q=2000$ for the $P$ waves \citep{Pasyanos&al.2009}, with travel times 
given by the ak135 reference Earth model \citep{Snoke2009}. }
\label{fig:P_vs_Rayleigh}
\end{figure}
From our calculations, we expect that $P$ waves will dominate the signal at large distances from the source. 
The exact location where $P$-wave levels overtakes Rayleigh-wave levels depends 
on the relative attenuation of the two types of waves. With a realistic $Q=2000$ for the $P$ waves, and 
$Q=400$ for the Rayleigh waves, figure \ref{fig:P_vs_Rayleigh} shows that it occurs at an epicentric angle of 40$^{\circ}$, which is a distance of 4400 km, 
consistent with the observations reported by \cite{Vinnik1973} using Kazakhstan array data.

\section{Conclusions}
We have shown how the same physical process, the interaction of ocean surface gravity wave (OSGW) trains, can produce 
a wide variety of noises, in the atmosphere, ocean and Earth's crust, that can be classified according to their 
horizontal phase speed. The slowest noises in the ocean are acoustic-gravity waves that dominate pressure records
 at depths less than about one tenth of the acoustic wavelength. These  acoustic-gravity waves cannot exist in the absence of 
OSGWs and are thus confined to the region of active wave forcing. Intermediate phase speeds correspond to 
Rayleigh waves that contain most of the energy of the seismic modes for distances less than about 4000~km from 
the source.  

We corrected previously published asymptotic behaviour for very long period 
noise ($T > 30$~s). In particular we find that the sources of this long period noise are attenuated on the continental shelves, consistent with previous 
studies of forced gravity wave motion. This finding supports a spatial distribution of these sources outside of the continental 
shelves, on the shelf breaks, which is consistent with data from \cite{Rhie&Romanowicz2006} or in deeper water, as reported by 
\cite{Nishida&Fukao2007}. 

The common source of all these noises should allow a verification of the source magnitude for seismic waves 
by near-surface measurements of pressure which is dominated by acoustic-gravity modes. 
In particular, the direct modelling of the acoustic-gravity modes can be compared to pressure 
measurements in depths less than a few hundreds of meters. 
Unlike the analysis of seismic noise 
\citep[e.g.][]{Ardhuin&al.2011}, which suffers from poorly known seismic 
propagation and attenuation factors, the acoustic-gravity wave attenuation over the water column 
can be predicted accurately \citep{Herbers&Guza1994} and thus pressure measurements in the upper ocean may provide a more 
quantitative verification of numerically modeled directional surface wave properties.  
In particular, as proposed by \cite{Cooper&Longuet-Higgins1951}, pressure measurements may provide 
a precise estimate of coastal reflection or wave scattering by currents, sea ice or other effects. 
Noise records from land-based or seafloor-mounted seismometers 
are more ambiguous because they integrate sources over a large area. 
Also, as discussed by \cite{Hasselmann1963c}, 
\cite{Abramovici1968}, and \cite{Latham&Sutton1966}, the variations in water depths and horizontal and vertical variations 
of properties in the Earth's crust can significantly modify noise properties in both the water column  and the crust.

F.A. is funded by ERC grant \#240009 ``IOWAGA'' with additional support from the U.S. National Ocean Partnership Program, 
under grant N00014-10-1-0383.  T. H. C. H. 
is supported by the U.S. Office of Naval Research Littoral Geosciences \& 
Optics Program and the U.S. National Oceanographic Partnership Program (NOPP).
Discussions with S. Webb and L. Mari{\'e} on the theoretical aspects are gratefully acknowledged. 

\appendix
\section{Coupling coefficients}\label{appA}
Using the coupling coefficient $D$ given by \citet[][eq. 4.3]{Hasselmann1962} 
for the velocity potentials, our coupling coefficient for the elevation amplitudes is  
\begin{eqnarray}
  D_z\left(\kb,s,\kpb,s'\right)& = & - \frac{ g^2 D \left(\kb,s,\kpb,s'\right)  }{ \ir s \sigma s'\sigma' \left(s \sigma+s' \sigma'\right) } 
  \nonumber \\
& =& \frac{g^2}{s \sigma s' \sigma'} \left\{  \left[\kb \bcdot \kpb - \frac{ \sigma^2 \sigma'^2}{g^2}  \right] \right. 
   +\left. \frac{0.5}{ \left(s \sigma+s' \sigma'\right)} 
\left(\frac{s \sigma k'^2}{ \cosh^2(k'h)}+\frac{s' \sigma' k^2}{ \cosh^2(k h)}\right)\right\} \nonumber \\
\label{Dz}
\end{eqnarray}

In the bottom pressure, the additional term arising from the orbital velocity has a coupling coefficient
\begin{eqnarray}
 D_{pb} \left(\kb,s,\kpb,s',z\right)  = \nonumber \\
    g^2 \frac{k k' \sinh[k(z+h)]\sinh[k'(z+h)] -  \kb \bcdot \kpb  \cosh[k(z+h)]\cosh[k'(z+h)]}
{2 s \sigma s' \sigma' \cosh(kh) \cosh(k'h)}. \nonumber \\  \label{Dpb}
\end{eqnarray}

The relationship with the coupling coefficient $C$ given by \citet[][their eq. 4]{Herbers&Guza1991} for the bottom pressure, expressed 
in meters of water, 
is given by solving eq. (\ref{eq:eq_phi_2}) for $\phi_2$, and then rewriting Bernoulli's equation (\ref{Bernoulli2}), as 
\begin{equation}
  \frac{p_2}{\rho_w} = \frac{\partial \phi_2}{\partial t} 
                         - \frac{1}{2}\left[
                                            \left|\bnabla \phi_1\right|^2
                                         +\left(\frac{\partial \phi_1}{\partial z}\right)^2\right]
\label{Bernoulli21}
\end{equation}
This gives, for $z=-h$,  
 \begin{eqnarray}
 C=-\frac{D_z(s\sigma + s' \sigma')^2 }{g \left[ g K \tanh(Kh) - (s \sigma + s' \sigma')^2 \right]} +  \frac{D_{pb}(z=-h)}{g}. 
 \end{eqnarray}
 
\clearpage
\section{Definition of symbols}\label{appB}
\begin{table}
  \centering
  \begin{tabular}{ccc}
  Symbol       & meaning  & where \\
\hline
   1 and 2  & indices denoting first and second order motions &   \\
   $\alpha_w$ & sound speed in water &   \\
   $\alpha_c$ & compression wave speed in crust &   \\
   $\alpha_a$ & sound speed in atmosphere &   \\
   $\beta$ & shear wave speed in crust &   \\
   $\Delta$ & angular distance  & \\
   $\delta$ & vertical ground displacement at the top of the crust, $\delta= \xi_z(z=-h)$& \\
   $\zeta$ & elevation of the sea surface &   \\
   $\lambda$ and $\mu$ & Lame elastic coefficients for the crust  &   \\
   $\lambda_O$ and $\phi_O$ & longitude and latitude of the observation location  &   \\
   $\lambda_S$ and $\phi_S$ & longitude and latitude of a noise source  &   \\
   $\rho$ & perturbation of density &   \\
   $\rho_w$ & mean water density &   \\
   $\sigma$ & radian frequency of surface gravity waves $\sigma=2\pi f$ &   \\
   $\tau$ &  stress tensor &   \\
   $\Theta$ & phase function of the seismic or acoustic waves &   \\
   $\phi$, $\phi_a$, $\phi_c$ & velocity potentials in the water, atmosphere, crust  &   \\
   $\psi$  & stream function in the crust  &   \\
   $\varphi$ & take-off angle for seismic body waves &   \\
    $\xi$ & displacement of particles &   \\
      $\omega_s$ & radian frequency of noise $\omega_s=2\pi f_s$ &   \\
$\omega_{s,j}^0$ & critical values of $\omega_s$ for which new modes appear when $\omega_s$ increases  &   \\
   \hline
\end{tabular}
  \caption{Table of arab and greek symbols}\label{table_symb}
\end{table}

\begin{table}
  \centering
  \begin{tabular}{ccc}
  Symbol       & meaning  & where \\
\hline
   $A, B$  &  amplitudes of $\phi_{c}$, $\psi$ & eq. (\ref{eq:phiAB}) \\
    $C, D$  &  amplitudes of up- and downward propagating components  of $\phi_2$  &  eq. (\ref{phi2CD})\\
   $C_g$  &  Group speed of surface gravity waves & eq. (\ref{eq:Cg}) \\
   $D_z$  & coupling coefficient for the surface elevations & eq. (\ref{Dz}) \\
   $a$ and $a'$ & amplitude of surface gravity waves \\
   $c_j$ & non-dimensional amplification factor for Rayleigh mode number $j$ \\
   $c_P$ and $c_S$ & non-dimensional amplification factor for P or S waves \\
   $f$ and $f'$ & frequency of surface gravity waves \\
   $f_s$  & acoustic or seismic frequency & \\
   $F_{p2,\mathrm{surf}}$  &spectral density of $\widehat{p}_{2,\mathrm{surf}}$ & \\
   $F_{\delta}$  &spectral density of $\delta$ & \\
   $G$  & surface pressure to bottom vertical displacement transfer function &  \\
  $g$  & apparent  gravity acceleration  & \\
     $h$  & water depth  & \\
   $I$& directional integral of the wave spectrum  & eq. (\ref{eq:I}) \\
   $j$ & Rayleigh wave mode number, counted from 0   &  \\

   $\kb$ and $\kpb$ & vector wavenumbers of surface gravity waves &\\
   $\Kb$& horizontal vector wavenumbers of acoustic or seismic waves &\\
   $l$, $l_a$, $m$ and $n$& vertical wavenumbers for $\phi_2$,$\phi_{2,a}$ , $\phi_{c}$, and $\psi$  &\\
   $p$  & pressure  & \\
   $ \widehat{p}_{2,\mathrm{surf}}$ & wave-induced forcing at the sea surface & eq. (\ref{P2hat}) \\
   $\widehat{p}_{2,\mathrm{B}}$  & Bernoulli head pressure & eq. (\ref{P2B}) \\
  $\widehat{p}_{2,\mathrm{bot}}$  &  wave-induced forcing at the bottom,  $\widehat{p}_{2,\mathrm{bot}}=\widehat{p}_{2,\mathrm{B}}(z=-h)$ & eq. (\ref{P2bot}) \\
   $M$  & matrix of the linear system of equations  &  \\
   $Q$  & seismic quality factor (i.e. damping coefficient) &  \\
   $q$ and $r$  & coefficients  & eq. (\ref{eq:syst3}) \\
   $R(K_{\max})$  & non-dimensional coefficient in microbarom source & eq. (\ref{RKmax}) \\
   $R_E$  & radius of the Earth &  \\
   $S_{DF}$  & Seismic source of Rayleigh waves &  \\
   $s$ and $s'$  & sign variables equal to -1 or 1  & \\
   $u$ and $w$  & horizontal and vertical velocity components  & \\
   $V$  & group speed of seismic waves  & \\
  $Z_{1,\kb}^{s}$  & complex amplitude of the linear surface elevation component $(\kb,s)$ &  \\
   \hline
\end{tabular}
  \caption{Table of roman notations}\label{table_symb3}
\end{table}

\clearpage

\end{document}